\documentclass[10pt,a4paper]{article}

\usepackage{a4wide}

\usepackage{float}
\usepackage{latexsym}
\usepackage{graphicx}
\usepackage{amsmath} 
\usepackage{amsfonts} 
\usepackage{color}
\usepackage{url}

\usepackage{amssymb,amsmath}
\usepackage{graphics,graphicx}
\DeclareGraphicsExtensions{.pdf,.eps}

\usepackage[english]{babel}
\usepackage{theorem}
\usepackage{bbm}
\usepackage{float}

\usepackage[round]{natbib}   
\usepackage{footnote}

\newtheorem{theorem}{Theorem}[section]
\newtheorem{corollary}[theorem]{Corollary}

{\theorembodyfont{\upshape}
\newtheorem{remark}[theorem]{Remark}
}

\newcommand{\Nmv}{N_{k\times d}(0,1_k\times 1_d)}

\newcommand{\R}{{\mathbb R}}

\newcommand{\N}{{\mathbb N}}

\newcommand{\cf}[1]{{\mathbbm 1}_{\{#1\}}} 
\newcommand{\cdata}{c_{data}} 
\newcommand{\cmodel}{c_{model}} 
\newcommand{\var}{\text{Var}} 
\newcommand{\corr}{\text{Corr}} 

\def\w#1{\mathop{:}\nolimits\!#1\!\mathop{:}\nolimits}

 \DeclareMathOperator*{\Var}{Var}
 \DeclareMathOperator*{\ccov}{Cov}

\newcommand{\BTB}{B^{T}B}
\newcommand{\Rd}{\R^{d}}

\newcommand{\RkN}{\R^{k\times N}}
\newcommand{\RNk}{\R^{N\times k}}
\newcommand{\RdN}{\R^{d\times N}}

\newcommand{\Rdk}{\R^{d\times k}}
\newcommand{\Rdd}{\R^{d\times d}}
\newcommand{\RNN}{\R^{N\times N}}
\newcommand{\spfivehundred}{{S\&{P}\;500}}

\newcommand{\E}{\mathbbm{E}}

\newcommand{\vep}{\varepsilon}

\newcommand{\email}[1]{E-mail: {\tt #1}}
\newcommand{\emailjani}{\email{jani.lukkarinen@helsinki.fi}}

\begin{document}

\title{Random selection of factors preserves the correlation structure in a linear factor model to a high degree}
%\shorttitle{Random selection of factors}

\selectlanguage{english}
%\begin{frontmatter}

\newcommand{\affilA}{Confederation of Finnish Industries EK, P.O. Box 30 (Etel\"{a}ranta 10), FI-00131 Helsinki, Finland.  \email{antti.tanskanen@ek.fi}}
\newcommand{\affilB}{University of Helsinki, Department of Mathematics and Statistics, P.O. Box 68, FI-00014 Helsingin yliopisto, Finland, \emailjani}
\newcommand{\affilC}{Varma Mutual Pension Insurance Company, P.O. Box 1, FI-00098 VARMA, Finland}

\author{%
Antti J. Tanskanen\thanks{\affilA},
Jani Lukkarinen\thanks{\affilB},
Kari Vatanen\thanks{\affilC}
}

\date{\today}

\maketitle 

\begin{abstract}
In a very high-dimensional vector space, two randomly-chosen vectors are almost orthogonal with high probability.
Starting from this observation, we develop a statistical factor model, the random factor model, in which factors are chosen at random based on the random projection method. 
Randomness of factors has the consequence that
covariance matrix is well preserved in a linear factor representation.
It also enables derivation of probabilistic bounds for the accuracy of the random factor representation of time-series, their cross-correlations and covariances. 
As an application, we analyze reproduction of time-series and their cross-correlation coefficients in the well-diversified Russell 3,000 equity index.
% \keywords{
% Factor modeling, equity markets, random variables \\
% %\ \\
% %JEL classification codes:  \\
% %SIB classification codes: 
% }
\end{abstract}

\section{Introduction}
\subsection{Vectors in a high-dimensional space}
Any two randomly-selected vectors in a high-dimensional vector space are likely almost orthogonal 
with respect to each other \citep{hecht1994context, kaski1998dimensionality}. This observation has relevance to time-series analysis, 
since a long time-series corresponds to a vector in a high-dimensional vector space and
orthogonality of vectors corresponds to uncorrelatedness of time-series.
Hence, time-series corresponding to the two randomly-selected vectors are almost uncorrelated.

When the length of the time-series increases, or equivalently, dimension of the vector space increases, the probability that two randomly-selected time-series are uncorrelated increases \citep{hecht1994context}.
If we select a set of, say, $n$ vectors randomly, these vectors are approximately orthogonal
to each other if the dimension of the space is sufficiently high. 
Then, high-dimensionality of the data may in some
cases even be an asset: in a high-dimensional space, almost any set of random vectors yields 
an almost uncorrelated set of factor time-series that can be used as a basis for a linear factor model. 

\subsection{Factor models}
Factor models are extensively used in financial applications to model asset returns (see, e.g., \cite{campbell1997econometrics})
and to decompose them to loadings of risk factors.
The two main types of factor models are fundamental factor models and statistical factor models. 
In a fundamental factor model, the aim is to find observable asset characteristics, 
e.g., financial ratios and macro-economic variables, capable of explaining the behavior of the market stock prices,
that are often extrinsic to the asset time-series.

The explanatory fundamental and economic variables can be highly correlated with each other, which may cause, e.g., multi-collinearity in a fundamental factor model.
Returns predicted by a fundamental factor model may then be more correlated than the observed returns, 
which is the main reason for the inclusion of specific risk components in a factor model.

Statistical factor models are a commonly-used alternative for fundamental factor models. 
In a statistical factor model, factors are extracted from asset returns. 
The principal component analysis (PCA; see, e.g., \cite{alexander2001market}) is an example of 
a statistical technique for finding factors from asset time-series. 

PCA works well when the analyzed time-series are highly correlated,
which may indicate the presence of a common driver. 
Applications of PCA include models of interest rate term structure, credit spreads, futures, and volatility forwards. 
In PCA, several principal components often have an intuitive financial interpretation. 
In ordered highly-correlated systems, the first principal component captures an almost parallel shift in all variables and is generally 
labeled the common trend component. 
The second principal component captures an almost linear tilt in the variables, 
while the higher order principal components capture changes that are quadratic, cubic and so forth (see, e.g., \cite{alexander2009market}). 
In the equity markets, the higher order principal components may often, but certainly not always, be interpreted as market movements
caused by different investment style tilts. 

\cite{stock2002forecasting} show that principal components found using PCA provide consistent estimators
of the true latent factors in the limit of both time and cross-sectional size go to infinity.
They extend consistent estimation of the classical factor model with non-correlated errors to approximate factor models with cross-correlated
and sectionally-correlated error terms.

Classical factor models include only a handful of factors. 
The best-known factor model in the literature is likely the Capital Asset Pricing Model (CAPM), which 
assumes that a single risk factor, the market, drives returns in a portfolio of assets  \citep{sharpe1964capital}. 
A number of factor models have extended this view (see, e.g.,  \cite{ross1976arbitrage, fama1993common}). 
Recent increase in computational power has enabled development of models with a large set of factors. 
Today, factor models are popular in market risk modeling, e.g., 
the Barra models  \citep{grinold2000active} depend on hand-picked market factors to explain behavior of the analyzed financial instruments. 

\cite{boivin2006more} demonstrate that there are situations in which the use of a larger number of time-series may
actually result a worse factor estimate than a smaller number of time-series.
A significant amount of recent literature has been devoted to address the issue
of consistent estimation under conditions where the number of time-series is large compared
to the length of time-series (e.g.,  \cite{bun2017cleaning,ledoit2012}).

\subsection{Choice of factors}
The choice of factors clearly influences the ability of 
a factor model to explain investment risk of a portfolio, in particular when the factor model
consists of only a few carefully-chosen factors. 
When the number of factors is large compared to the number of time-series analyzed, 
it may not much matter which factors are chosen
as long as the factors span a sufficiently large sample space. 
Even then, relative importance of factors is often of interest in risk management.
However, it is not clear how well an arbitrary set of factors would enable 
analysis - or at least description - of the risk. 
This is the issue that we analyze in this study: 
take a random set of factors and see whether it enables reproduction of the data
and its interdependencies.

A good starting point for developing a random factor model is 
the random projection method (see, e.g, \cite{bingham2001random, vempala2005random})
that consists of a projection of data to a lower-dimensional space by a random matrix.
The random projection method has been used, e.g., to reduce the complexity of 
the data for classification purposes  (\cite{kohonen2000self}), 
for structure-preserving perturbation of confidential data in scientific applications  \citep{liu2006random},
for data compression \citep{bingham2001random}, for compression of images  \citep{amador2007random},
and in the design of approximation algorithms  \citep{Blum2006}.
The random projection allows one to reduce dimensionality of the investigated problem, often substantially, 
while preserving the structure of the problem.  

From the econometric point of view, a random projection can be viewed as a projection of time-series to
a collection of almost non-autocorrelated factors that are also expected to be non-cross-correlated,
as will be seen in Section 2.

\subsection{Correlation structure and random matrices}
Analysis of risk in an investment portfolio requires that risks of 
individual instruments are combined into the risk of the portfolio.
This can be accomplished using dependence structures, e.g., correlation matrix. 
It is no coincidence that dependence structures of financial time-series 
are central in modern investment theory  (e.g., \cite{markowitz1952portfolio}). 

The recent explosion of available data has brought new issues with time-series analysis (e.g., \cite{bun2017cleaning}).
To get around these issues, new methods of analysis are needed. One such tool is the random matrix theory 
(e.g.,  \cite{edelman2005random}).
It is typical in the random matrix theory that things get less complex when dimension of the problem increases. 

A number of studies have applied random matrix theory to analyze correlations.
The Tracy-Widom distribution \citep{tracy1996orthogonal} gives distribution of the largest 
normalized eigenvalue in Gaussian Orthogonal Ensemble.
There are also versions of the theorem for other random matrix ensembles \citep{tracy1994level, tracy1996orthogonal}.
The Tracy-Widom distribution is an example of universality found in the random matrix theory. 
Another important example of universality in random matrix theory is the Marchenko-Pastur law  \citep{marchenko1967distribution}, which 
describes the asymptotic distribution of eigenvalues of a covariance matrix.

\cite{laloux1999noise} argue that most of the eigenvalue spectrum of correlation matrix 
for the Standard \& Poor's 500 Index (\spfivehundred) equities can be described
as a product of noise, only about 20 eigenvalues of 500 total are informative. 
In more detail, \cite{laloux1999noise} demonstrated that the bulk of 
the power spectrum of the \spfivehundred\ returns is indistinguishable from 
that produced by Gaussian Orthogonal Ensemble, in which the asymptotic eigenvalue distribution is
given by the Marchenko-Pastur law  \citep{marchenko1967distribution}.
Since Gaussian Orthogonal Ensemble consists of random matrices, the bulk of power spectrum can be explained as a product of noise. 

The informative eigenvalues may correspond to the market movements and sectors, 
but most of the eigenvalue spectrum does not correspond to linear factors. 
This suggests that noise has a significant role in the description of dependence structures in financial data.

\cite{plerou1999universal} show that the most eigenvalues in the spectrum of the cross-correlation matrix of stock
price changes agree surprisingly well with universal predictions of random matrix theory. 
\cite{malevergne2004collective} construct a model with a spectrum that exhibits
the features found in \spfivehundred\ spectrum: a few large eigenvalues and a bulk part. 
\cite{malevergne2004collective} also point out the chicken-and-egg problem associated with factors:
factors exist because stocks are correlated; stocks are correlated because of a common factor impacting them.
They argue that the apparent presence of factors is a consequence of the collective, 
bottom-up effect of the underlying time-series.

\cite{ledoit2004honey} show that a naive use of the sample covariance matrix has drawbacks,
and that shrinkage of the sample covariance matrix toward a structured or a constant matrix 
reduces extremal estimation errors. In a subsequent work, \cite{ledoit2012} use Marchenko-Pastur law   
to improve the results further. 

\cite{bun2017cleaning} give a concise review of the random matrix theory and then move on to show how it can be used to alleviate
the issues of "Big data". 
In particular, they describe the issues of correlation matrix estimation when there are more time-series than data points in each time-series.
They also suggest remedies based on the the random matrix theory that  can be used to improve empirical estimates
of correlation matrices. 

Given the above examples, it is clear that the random matrix theory has already contributed to
the analysis of financial time-series. 
We make use of random matrices to develop a factor model and to analyze the properties of correlation
matrix preservation in a generic linear factor model.

\subsection{This study}
It is largely an open question whether and how well randomly-chosen factors can be used to
describe a large data set. This is the issue we approach in this study. 
For this purpose we develop a factor model based on randomly-chosen factor time-series, the random factor model. 
We show that randomness of factors has certain desirable properties, 
such as well-defined probabilistic limits on the accuracy of the factor representation. 
In addition, randomly-chosen factors are almost orthogonal with high probability, and
with a proper normalization, they are expected to be orthonormal.
We also show how the random factor model converges
toward the modeled data when the number of factors increases,
and that a random factor model preserves pair-wise correlations well with high probability.

The article is structured as follows. In Section 2, 
we develop the random factor model based on the random projection method
and derive theoretical results describing the model (more details can be found from Appendix A).
For example, we show that randomness of factors enables derivation of theoretical
results on the accuracy of the model.

As an application of the random factor model, 
Section 3 provides an analysis of the correlation matrix of Russell 3,000 equity index using the factor models described in Section 2.  
We analyze the ability of random factor models to reproduce equity log-return time-series and their correlations and covariances. 
The reproduction of data in a random factor model is compared with a reproduction obtained using principal component analysis 
both at the individual time-series level and at the dependence structure level. 

In Section 4, we compare different random factor models and show that the results, or rather
their accuracy, are quite universal.
In Section 5, we discuss the results and their possible implications.

In Appendix A, we prove a version of the Johnson-Lindenstrauss-like theorem appropriate for the random factor model. 
It gives probabilistic bounds on the accuracy of the correlation and covariance preservation in a 
random factor model.

\section{Random factor model}
\subsection{Notations}
Let $d$ observations of time-series $Z: \N\to\R$  be viewed as a vector in $d$-dimensional space $\Rd$, 
where each observation of the time-series corresponds to one coordinate of the vector. 
The set of $N$ such time-series  can be packed into matrix $X\in \RdN$, in which observations are in columns. 
We assume that the time-series data has been preprocessed, so that each 
time-series is averaged to zero, that is, $\sum_m X_{m b}=0$ for each $b=1,...,N$.
We employ sample statistics in this study. Definitions for mean $\mu$, variance $\sigma^2$ and covariance $C$ are
\begin{align}
    \mu_x={\frac{1}{d}}\sum_{m=1}^dx_m, \qquad
    C_{x,y}={\frac{1}{d-1}}\sum_{m=1}^d(x_m-\mu_x)(y_m-\mu_y), \qquad
    \sigma^2_x=C_{x,x}, \qquad
\end{align}
where $x,y\in\Rd$.
The central parameters used in this study are summarized in Table \ref{table:1}. 

\begin{table} %[h!]
  \begin{center}
    \begin{tabular}{| l | c |}
    \hline
    Parameter & Description \\
    \hline
    $d$ & Number of data points in each time-series \\
    $N$ & Number of time-series, e.g., equities \\
    $k$ & Number of factors \\
    \hline
    \end{tabular}
  \end{center}
  \caption{The central parameters used.}
  \label{table:1}
\end{table}

\subsection{Linear factor models}
A linear factor model describes the target data set as a loading-weighted sum of factors   (e.g., \cite{mcneil2015quantitative}). 
Let $F=[F_1\, F_2\, ...\, F_k]\in \R^{d\times k}$ contain $d$ observations of the factors $j=1,2,\ldots,k$, $F_j\in \R^{d\times 1}$. 
Then time-series $X_b\in \R^{d\times 1}$, where $X_b$ is $b$:th column of matrix $X$, $b=1,...,N$, can be represented
as a sum of products of factor loadings $L_{bj}\in \R$ and factors $F_j$, that is,
\begin{equation}\label{eq:1}
	X_b=\sum_{j=1}^k L_{bj}F_j+\epsilon_b,
\end{equation}
where $\epsilon_b$ is an idiosyncratic risk component.  
Since we collect the observations into columns of $X$ and $F$, the formula (\ref{eq:1}) is written in a matrix form 
as $X=F L^T+\epsilon$.

Factors in $F$ may or may not be directly observable in the market data. 
For observed factor time-series, it suffices to project the data to factors to get loadings.
For unobserved factor time-series, some method such as PCA or some other optimization-like method
is required to find the factors and their loadings.

\subsection{Random projection}\label{sec:randomproj}
Random factors are here chosen using the random projection method. 
The key idea of random projection is based on the Johnson-Lindenstrauss lemma  \citep{johnson1984extensions,bingham2001random}: 
if points in a high-dimensional space are projected onto a randomly selected subspace of suitably high dimension, 
then the distances between the points are approximately preserved. 
A suitably high-dimensional subspace has dimension proportional to $\log(N)/\varepsilon$, 
where $N$ is the number of time-series and $\varepsilon$ the desired accuracy  \citep{dasgupta2003elementary}.

Random projection $Q: \RdN\to\RkN$ of matrix $D\in\RdN$ 
is a mapping defined by $Q(D)=BD$, where $k,d,N\in \N$.
Here matrix $B$ is realization $B(\omega)$ of a random variable-valued $k\times d$ 
matrix\footnote{From this point on, we will not differentiate between 
random variables and their realizations and tacitly assume that the distinction can be inferred from the context.}.

A large variety of probability distributions can be used to construct projection matrix $B$ (more on this in Section 3.3).
The most obvious choice is to assume that matrix $B$ is taken from the matrix-variate normal distribution with independent entries, 
that is, from $\Nmv$  \citep{gupta1999matrix}.
Then each element is $N(0,1)$-distributed and independent of other elements.

\subsection{Random factor model}\label{sec:randomfactormodel}

\subsubsection{Definition and properties}

We define the random factor model (RFM) for data set $X\in\RdN$ via 
a projection\footnote{Strictly speaking, the matrix $P$ is not a projection matrix since it typically does not satisfy the equality $P^2=P$.  However, 
since its range is a lower dimensional subspace, we use the term ``projection'' also to describe $P$, in analogy with the definition of the term ``random projection'' in Section \ref{sec:randomproj}.}
$P:\RdN\to\RdN$, 
\begin{equation}\label{eq:2}
 PX=aB^TBX,
\end{equation}
where $B\in\Nmv$ is a $k\times d$-dimensional random variable, elements of which are
independent and normally distributed, and $a>0$ is a normalization constant. 
Mapping (\ref{eq:2}) can be interpreted as a linear factor model by setting
\begin{equation}\label{eq:4}
L=\frac{a}{a'} X^T B^T,\\
\quad F=a' B^T\, ,
\end{equation}
where $a'>0$ is a constant related to factor normalization, as
discussed in Section \ref{sec:orthogonality}.
Then $F\in\Rdk$ behaves as a matrix of 
$k$ $d$-dimensional factor time-series. 
It is worth stressing that matrix $F$ consists of random time-series that 
in no way depend on the data.
$L\in\RNk$ is a matrix of $k$ factor loadings for the $N$ time-series.

Projection $P$ can be factored as
\begin{equation}\label{eq:3}
PX=F L^T.
\end{equation}
Defining $\epsilon^*=X-PX$ yields an approximate factorization 
\begin{equation}\label{eq:reps}
X=F L^T+\epsilon^*
\end{equation}
for data matrix X. We will analyze equation (\ref{eq:reps}), and in particular error term $\epsilon^*$, further in the following.
Equation (\ref{eq:reps}) shows that data matrix $X$ can be approximately decomposed 
to a product of two components.

As an aside, let us mention that we could equally well have 
considered a random projection in the equity direction instead of the above time-series direction.
This can be accomplished using a matrix $Q=a R^T R$, where $a>0$ and $R\in\RkN$ is a random matrix,
and then considering $XQ$ as the projected matrix.  This naturally leads to a factor model interpretation
with a loading matrix $\frac{a}{a'} R^T$ and a factor matrix $a' X R^T$.  In analogy to the earlier terminology, this model could be called {\em random loading model}.  The properties proven later for the random projection $P$ then immediately carry over to the projection $Q$, one merely needs to replace ``$d$'' with ``$N$'' in all of the results.

However, from the point of view of the time series, the two projection methods $PX$ and $XQ$ could behave differently.
For instance, if there are more pronounced correlations between different equities at a fixed time than between the same equity at two different times, then one would expect to need larger values of $k$ in the projection $XQ$ than in the projection $PX$ to reach the same level of accuracy in the approximation.  It is also possible to apply both random projections simultaneously and study $PXQ$ instead of $PX$ or $XQ$.  This double-sided projection would still have properties very similar to the one-sided projections, as long as the random matrices $B$ and $R$ are chosen independently of each other.   
Since the three alternatives are on a technical level very similar, we focus only on the choice $PX$ in the following.

As the next step, we need to find a suitable constant $a$ so that standard deviation, covariance and 
the expected value of the data are preserved, if possible. 
Under these conditions, $\epsilon^*$ should be close to zero.
Different choices of $a$ yield slightly different properties for the RFM,
but it turns out that we cannot satisfy all these requirements at the same time if we base matrix $B$ 
on the normal distribution\footnote{For example, it follows from the results proven in the Appendix that, when 
$a=1/k$, representation (\ref{eq:1}) is exact on average, in the sense that $\E[PX]=X$. 
However, this representation over-estimates the sample variance $\sigma_x^2$ of 
a time-series $x\in \R^{d\times 1}$, since then $\E[\sigma_{Px}^2]=(1+d/k)\sigma_x^2$.  Hence, the 
asset returns would fluctuate too much in this normalization in the typical regime where $k\ll d$.
In addition, although the projection would then produce the correct time-series on average, 
the actual values are dominated by fluctuations: the 
standard deviation of $(Px)_m$ is at least $\sigma_x \sqrt{d/k}$ and thus one given sample of the random factors is 
unlikely to be a useful representation of the data, unless $k$ is at least comparable to $d$.}.

Here we concentrate on preserving the covariance matrix $C_{x,y}$ in the projection. 
Then normalization constant $a>0$ must be such that expectation with respect to $\Nmv$ is preserved, that is,
\begin{equation}\label{eq:5}
\E[C_{Px,Py}]=C_{x,y}
\end{equation}
for any zero-mean vectors $x, y\in \R^{d\times 1}$.  
It is worth stressing that the expectation in equation (\ref{eq:5}) is taken over random factor models, not over time-series $x$ and $y$.
Theorem \ref{th:main} in the Appendix shows that this is possible but only if we choose $a=1/\sqrt{k(k+d)}$. 
Let $a$ have this value from this point on.

The expected covariance between time-series $x$ and $y$ is then preserved, regardless of the number of factors used. 
Since $\E[\sigma^2_{Px}]=\E[C_{Px,Px}]=C_{x,x}=\sigma^2_{x}$, 
our choice of $a$ also preserves time-series variance. 
This result shows that an RFM is expected to fulfill the consistency requirement of variance,
that is, it shows that 
\begin{equation}\label{eq:consistency}
 \lim_{d\to\infty}\E[\sigma^2_{Px,d}]=\lim_{d\to\infty}\sigma^2_{x,d}=\sigma^2_{x,\text{pop}},
\end{equation}
where $\sigma^2_{x,\text{pop}}$ is the population variance and $\sigma^2_{x,d}$ is the sample variance in dimension $d$. 
An application of the Jensen's inequality implies that
$\E[\sigma_{Px}]\leq\sigma_x$, that is, volatility is not over-estimated.

Representation (\ref{eq:reps}) always preserves the average of a zero-mean vector $x\in\Rd$, that is, $\E[\mu_{Px}]=0$. 
In contrast, the $m$:th observation $x_m$ of time-series $x$ has an expectation $\E[(Px)_m]=\sqrt{k/(k+d)}x_m$
and a variance $(x_m^2+(d-1) \sigma_x^2)/(d+k)$.
For a small number of factors, mapping to $(Px)_m$ will on average underestimate the original value $x_m$ since $\sqrt{k/(k+d)}<1$. 
In the limit of large number of factors\footnote{In the RFM, 
the number of factors $k$ is not limited either by $N$ or by $d$.}, $(Px)_m$ approaches $x_m$, since
\begin{equation}\label{eq:7}
 \lim_{k\to\infty}\E[(Px)_m]=\lim_{k\to\infty}\sqrt{k/(k+d)}x_m=x_m\, ,
\end{equation}
and the standard deviation of $(Px)_m$ is $O\left(\sqrt{d/(d+k)}\right)$ and thus goes to zero when $k\to \infty$.
Hence, a RFM reproduces any vector $x\in\R^n$ component-by-component 
in the limit of large number of number of factors, for $k\gg d$. 
Thus $\epsilon$ of equation (\ref{eq:reps}) approaches zero when the number of factors increases.

The RFM is expected to reproduce mean, variance and covariance of 
time-series $x$. 
Component-wisely, the random factor model is expected to converge 
to the observed component values in the limit of large number of factors.

\subsubsection{Covariance preservation}

Equation (\ref{eq:5}) does not state that each RFM always preserves the covariance matrix. 
Nevertheless, it is reasonable to assume that an RFM approximately preserves the covariance matrix.
Next we will analyze how well an RFM will typically preserve the covariance matrix.

But first, it is worth recalling that Johnson-Lindenstrauss theorem  \citep{johnson1984extensions, dasgupta2003elementary, matouvsek2008variants}
gives probabilistic bounds for the accuracy of distance preservation in the random projection. 
A number of versions of Johnson-Lindestrauss theorem have been proven, however,
in all versions known to us, it is assumed that random variables have zero expectation.

Matrix $B^TB\in\Rdd$ is a singular Wishart matrix (also known as an anti-Wishart matrix), which has non-zero expectation,
$d-k$ zero eigenvalues and $k$ non-zero eigenvalues.
Since matrix $B^TB$ has non-zero expectation, it was not {\em a priori\/} clear if a Johnson-Lindenstrauss type theorem holds.
Theorem \ref{th:main} proven in the Appendix fills this gap for the present type of anti-Wishart matrices, and it
also contains a detailed derivation of the above expectation values for an arbitrary value of the scaling parameter $a$.

We have collected in Corollary \ref{th:coroll} the corresponding results for the choice which 
preserves the sample covariance matrices in expectation, for $a=1/\sqrt{k(k+d)}$.  
The precise control of fluctuations in the covariance estimates requires nontrivial
combinatorial computations, given in the Appendix.  As proven in Corollary  \ref{th:coroll},
for every $b>0$ and non-random vectors $u,v\in \R^d$, with $\mu_u=0=\mu_v$, we have\footnote{Since 
our proof is based on the Chebyshev inequality, there could still be room for improvement in the estimate. 
Also, as noted in Remark \ref{th:constantdisc} after the proof, 
the prefactor $8$ in Equation (\ref{eq:t1}) is not always optimal, and it could be reduced to $2$ in the regime
$d\gg k$.}
\begin{equation}\label{eq:t1}
  \mathbbm{P}\!\left[|C_{Pu,Pv}-C_{u,v}|\ge b\right]\le \frac{8}{k b^2} \sigma_u^2\sigma_v^2 \, .
\end{equation}
Inequality (\ref{eq:t1}) gives bounds on the accuracy of covariance preservation for an arbitrary random factor model. Here probability is taken with respect to an ensemble of random factor models.

Hence, if $\sigma_u^2,\sigma_v^2\le 1$, the probability that covariance of vectors $u$ and $v$
is preserved in a random factor model more accurately than bound $b$ is at least $1-8/(kb^2)$, 
where $k$ is the number of factors.  In general, we can set $b=\varepsilon \sigma_u\sigma_v$, 
with $\varepsilon>0$, and also conclude that the accuracy, relative to the sample variance scale $\sigma_u\sigma_v$,
is at least $\varepsilon$ with a probability of at least $1-8/(k\varepsilon^2)$.  For the bound to be informative,
it is necessary that $\varepsilon> k^{-1/2}$.

The error in the covariance estimate decreases at least inversely with the number of factors in almost any random factor model. 
Given a sufficient number of factors, covariance of 
any two time-series can be approximated with an arbitrarily high accuracy using an RFM.
This and the fact that random factors are in no way fitted to the data
suggest that the typical accuracy of an RFM depends mainly on
the number of factors $k$. 

Corollary \ref{th:coroll} also gives a bound on how accurately correlation between projected vectors is preserved.
Correlation $\corr(u,v)$ coincides with covariance $C_{u,v}$ when $\sigma_u=\sigma_v=1$.
Then inequality (\ref{eq:t1}) gives a lower bound
on how well $C_{Pu,Pv}$ approximates correlation between $u$ and $v$. 

These results can be summarized as a statement about the projected matrix $PX$ as follows:
\begin{equation}\label{eq:Xaccurary}
  \mathbbm{P}\!\left[\frac{1}{\sigma_{X_{b}}\sigma_{X_c}}
   |C_{(PX)_{b},(PX)_c}-C_{X_{b},X_{c}}|< \vep\right]\ge 1-\frac{8}{k \vep^2}\, ,
\end{equation}
valid for any $b,c= 1,2,\ldots,N$ and $\vep>0$.

\subsubsection{Almost orthogonality}\label{sec:orthogonality}

Orthogonality is a desirable property of a factor set. 
An orthogonalization procedure can be used to obtain an orthogonal factor set, but orthogonalization  is
computationally expensive. 
Fortunately, orthogonalization is not a necessary step in the RFM. 

Given any two random factors (as defined above), 
their inner product is expected to be orthogonal, that is,
\begin{equation}\label{eq:11}
\E\left[\sum_{m=1}^d F_{mj}F_{mj'}\right]=(a')^2 
\sum_m\E\left[B_{j'm}B_{jm}\right]=(a')^2 d \delta_{j',j}.
\end{equation}
%\nb{since $\sum_j F_{cj}F_{bj} = a (Pu)_c$ for the vector $u_n=\delta_{n,b}$.}
This shows that with the choice $a'=1/\sqrt{d}$, the factors
$F_{j'}$ and $F_j$ are expected to be orthonormal as
a consequence of the properties of normally distributed random variables.

Using Theorem A.1 
we can also compute the variance of the inner product.  This yields
\begin{equation}\label{eq:12}
\var\left[\sum_{m=1}^d F_{mj}F_{mj'}\right] = \frac{1}{d} ( \delta_{j',j} + 1 ) \le 
\frac{2}{d}.
\end{equation}
Higher cumulants approach zero even more rapidly, as can be seen by analyzing the cumulant generating function
\begin{equation}
\ln \E\left[e^{\lambda\sum_{m=1}^d F_{mj}F_{mj'}}\right]=\begin{cases}
-\frac{d}{2}\ln(1-2\frac{\lambda}{d}), \quad \text{when $j=j'$,}\\
-\frac{d}{2}\ln(1-\frac{\lambda^2}{d^2}), \quad \text{otherwise,}
\end{cases}
\end{equation}
and its series expansion in $\lambda$. When $j=j'$, $n$th cumulant is of order $O(d^{1-n})$. When $j\neq j'$,
cumulants are of order $O(d^{1-n})$ for even $n$ and zero otherwise.
Convergence to the Normal distribution in the limit of large $d$
then follows by the standard arguments (e.g., \cite{billingsley1995probability}).

Inner product matrix is approximately distributed as
\begin{equation}\label{eq:13}
\sum_{m=1}^d F_{mj}F_{mj'} \sim\begin{cases}
N(1,\sqrt{2/d}), \text{ when } j'=j,\\
N(0,\sqrt{1/d}), \text{ otherwise}.\\
\end{cases}
\end{equation}
When $d$ is large ($\gg 1000$), standard deviation is only a fraction of the expectation for diagonal elements.
Fluctuations around zero are small for non-diagonal elements.
The cumulant expansion shows that the factors are almost orthonormal even at a relatively low dimension. 

In addition, the factors are on average orthogonal to the error term $\epsilon^*$:
since $\epsilon^*$ is an even polynomial of $B$:s and $F_j$ is linear in them,
we have $\E\left[\sum_{m=1}^d F_{mj} \epsilon^*_{mb}\right]=0$ for all $j=1,2,\ldots,k$ and $b=1,2,\ldots,N$.

\subsection{Principal component analysis}

PCA is a well-known technique which uses a linear transformation to form a simplified data set retaining the characteristics of the original data set (see, e.g., \cite{johnson2014applied}). 
In investment risk measurement, PCA is used to explain the covariance structure of a set of market variables through a few linear combinations of these variables. 
The general objectives of using PCA are to reduce the dimensions of covariance matrices and to find the main risk factors. 
The risk factors can then be used to analyze, e.g., the investment risk sources of a portfolio, or to predict how the value of the portfolio will develop.

Projection to principal components is most directly obtained using the singular value decomposition \citep{golub2012matrix}. Given data matrix $X\in \RdN$, SVD decomposes it as $X=P_LDP_R^T$, where $P_L\in \Rdd$ is matrix of left singular vectors, $P_R\in \RNN$ is matrix of right singular vectors, and $D\in \RdN$ is the rectangular diagonal matrix 
of singular values. PCA-based factor representation of $X$ is given by $X=F L^T$, where $L=P_R\in \RNN$ gives the factor loading matrix and $F=P_LD \in \RdN$ defines the factors of the $N$ equities.
When reducing the dimensions of the original dataset, the first $k$ principal components with the largest eigenvalues are chosen to represent the original dataset. 
This yields an approximation of the data matrix using a subset of factors. 
A $k$-factor approximation of matrix $X$ is given by $F^{(k)}(L^{(k)})^T$, where $L^{(k)}\in\RNk$ contains the first $k$ factor loadings, $F^{(k)}\in \Rdk$ contains the components of the first $k$ factors from $P_L D$. 
It can be shown that in the mean-error sense, PCA gives the best linear $k$-factor approximation to matrix $X$ (e.g.,  \cite{reris2005, eckart1936approximation}). 
Principal components correspond to directions along which there is most variation in the data. However, there are no guarantees that pair-wise distances are preserved in PCA.

PCA yields the relative importance of the most important risk sources (defined in factor matrix $F$) in an investment portfolio. 
The relative importance of risk factors is shown by the size of eigenvalues. 
The eigenvectors with highest eigenvalues correspond to the most important risk factors. 
Loadings then tell how much investment instruments depend on these factors. 

Nevertheless, it should be stated that PCA aims to capture total
variation, not correlations \citep{johnson2014applied}.

\subsection{Comparison of factor models}
Despite appearance, the RFM and PCA share many features. 
In both models, the data can be represented as $F L^T$, 
where $L$ contains $k$ factor loadings and $F$ defines $k$ factor time-series with $d$ observations.  
In PCA, the most important eigenvectors are found
by choosing the largest eigenvalues. No such ordering is available for random vectors. 
A random vector is essentially as good as the next random vector as a factor.

The RFM has an {\em almost} orthonormal factors, while PCA yields strictly orthonormal factors. 
After finding the factors, both the RFM and PCA project the data to these vectors. 
The ways in which the RFM and PCA end up with representations of the data matrix are quite different: 
in PCA, data is projected along principal components (factors) and only the desired set of these projections (loadings) are kept. 
In the RFM, the data is projected along the random factors. 
The main difference is in the way that factors are chosen.

PCA requires $O(d^2N)+O(d^3)$ operations, while the RFM requires $O(kdN)$ operations, given the factor time-series. 
Since that the number of factors is typically significantly smaller than dimension of data, 
the RFM is computationally much more efficient than PCA. 

We do not aim at proving the supremacy of the RFM over PCA. 
We rather use PCA as a yardstick against which the RFM is compared. 
It is worth remembering that there is no fitting to data in the RFM, so one could 
reasonably expect that PCA would surpass RFM in every respect in data experiments.

\section{Application to the Russell 3,000 equity index}
The Russell 3,000 index (ticker RAY in Bloomberg) measures the performance of the 3,000 largest US companies.
The index represents about 98 percent of the investable US equity market.
Here, we investigate how well a random factor model 
reproduces log-returns of the Russell 3,000 equities and their cross-correlations, 
and compare the results to those obtained using PCA. 

For our analyses, we employ daily log-returns of Russell 3,000 equities from 2000-01-03 to 2013-02-20 (in total 3,305 observations). 
This interval contains several phases of the business cycle and certain special events, e.g., the crash of September 2008.
Of the 3,000 constituent time-series in the index, we used a subset of 1,591 time-series with continuous daily data
covering the entire period. 
To apply the analysis methods, the data is normalized by subtracting mean of each return 
time-series and by dividing by its standard deviation. 

\begin{figure} 
\includegraphics[width=15cm]{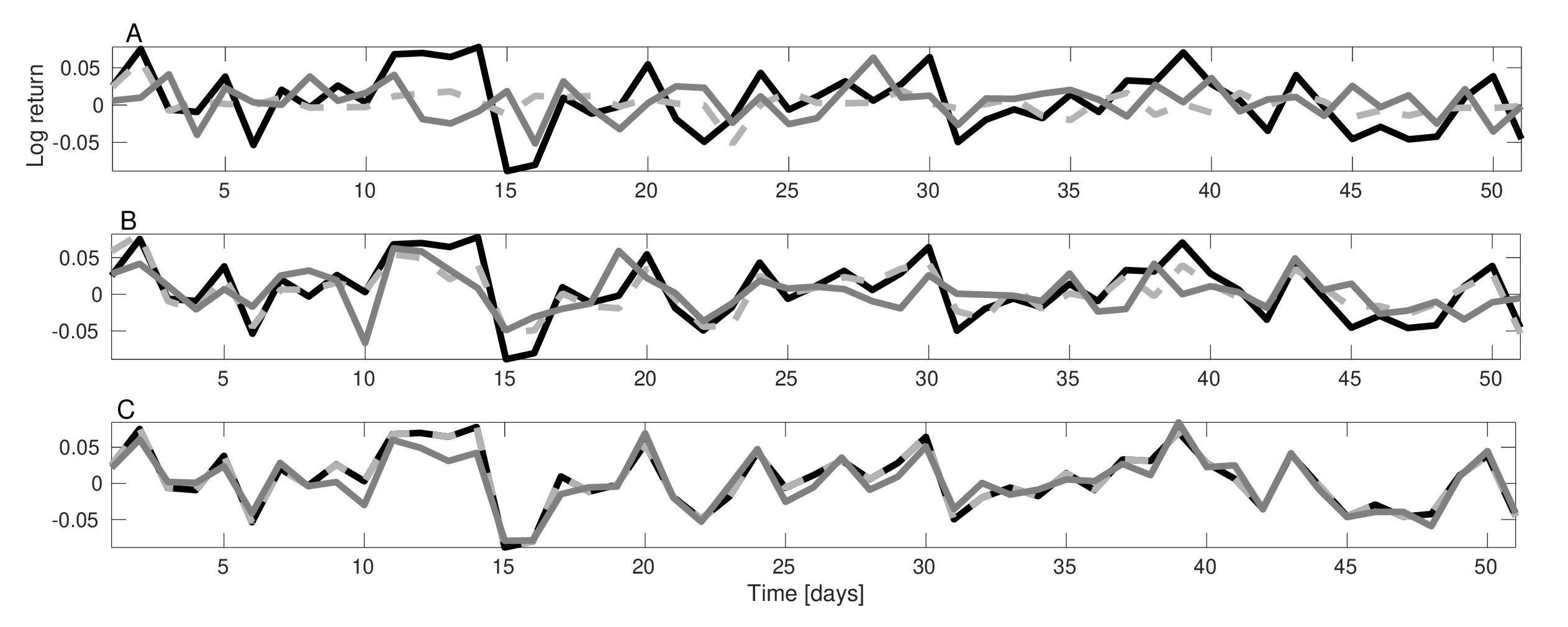}
\caption{An example reproduction of logarithmic return time-series using the random factor model (dark grey curves) and PCA (dashed light grey curves)
compared with the data (solid black curves) using 
(A) 10 factors, (B) 100 factors, and (C) 500 factors. The data is normalized to have zero average return, as describes in the text.}
\label{fig:fig1}
\end{figure}

\subsection{Reproduction of time-series}

Fig.\ 1 provides three examples of time-series reproduction using the RFM and PCA.
The RFM (grey solid curve in Fig.\ 1) provides a 
good reproduction of the single time-series even with a low number of factors. 
The accuracy of the reproduction improves with the number of factors: the agreement
of the RFM and the data is very good with 500 factors. 

The number of factors is not limited to the number of time-series in the RFM,
since the random factors do not necessarily span the entire space in which 
time-series may have values. Only in the limit of large number of factors
is the entire space covered.

\begin{figure} 
\includegraphics[width=15cm]{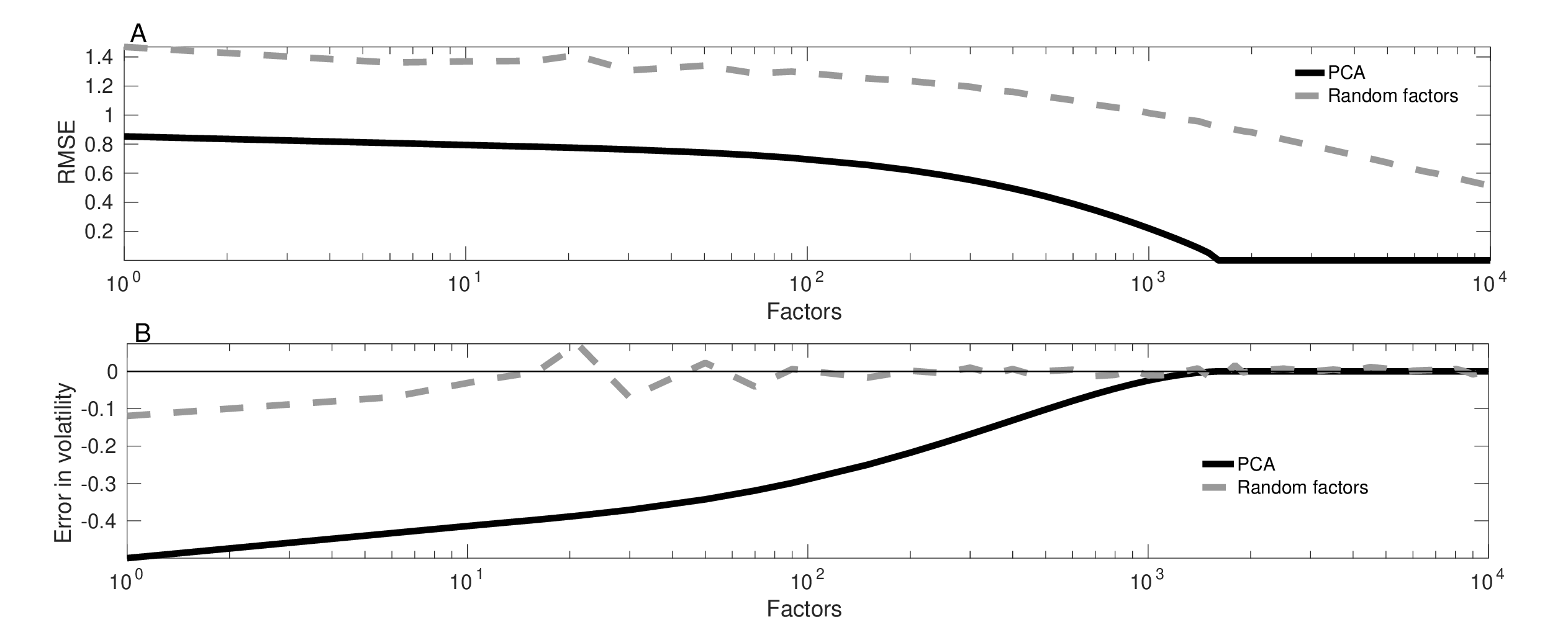}
\caption{Accuracy of time-series representations. 
(A) Error in time-series reproduction using the random factor model (dashed gray curve) and PCA (black solid curve) measured by RMSE. 
Curves are shown as functions of the number of factors.
(B) Error in reproduction of time-series volatility using the random factor model (dashed gray curve) and PCA (black solid curve) 
as a function of the number of factors. Errors are relative to the volatility of the time-series due to normalization.
}
\label{fig:fig2}
\end{figure}

Both PCA and the RFM provide good reproductions of the data (Fig.\ 1),
however, there are deviations from the data in each reproduction. 
In the root mean square error (RMSE) sense, 
PCA gives a better reproduction of the time-series than the RFM  (Fig.\ 2A).
RMSE in the reproduction of the entire data set is 0.79 in PCA vs 1.37 in the RFM with 10 factors (Fig.\ 1A). 

\subsection{Volatility}

The RFM reproduces volatility of the time-series almost exactly even with a small number of factors, 
while in PCA volatility estimates improve pronouncedly with more factors (Fig.\ 2B). 
Since volatility of each time-series is normalized to 1 separately,
accuracy of volatility reproduction is relative to volatilities of the underlying time-series in Fig.\ 2B.

In the RFM, error in volatility\footnote{Error is here defined as the difference $\cmodel-c_{data}$ between 
correlation $\cmodel$ estimated from the modeled data 
and correlation $c_{data}$ computed from the original data.} is about 3.1 percent of volatility with ten factors.
In PCA, error is about 41.7 percent of volatility with ten factors. 
Accuracy increases until 1,000 factors is reached, after which essentially no error is observed in PCA.
While the RFM reproduces the overall volatility of the equity time-series faithfully, 
it does not capture time-dependence of volatility particularly well (data not shown).

\begin{figure} 
\includegraphics[width=15cm]{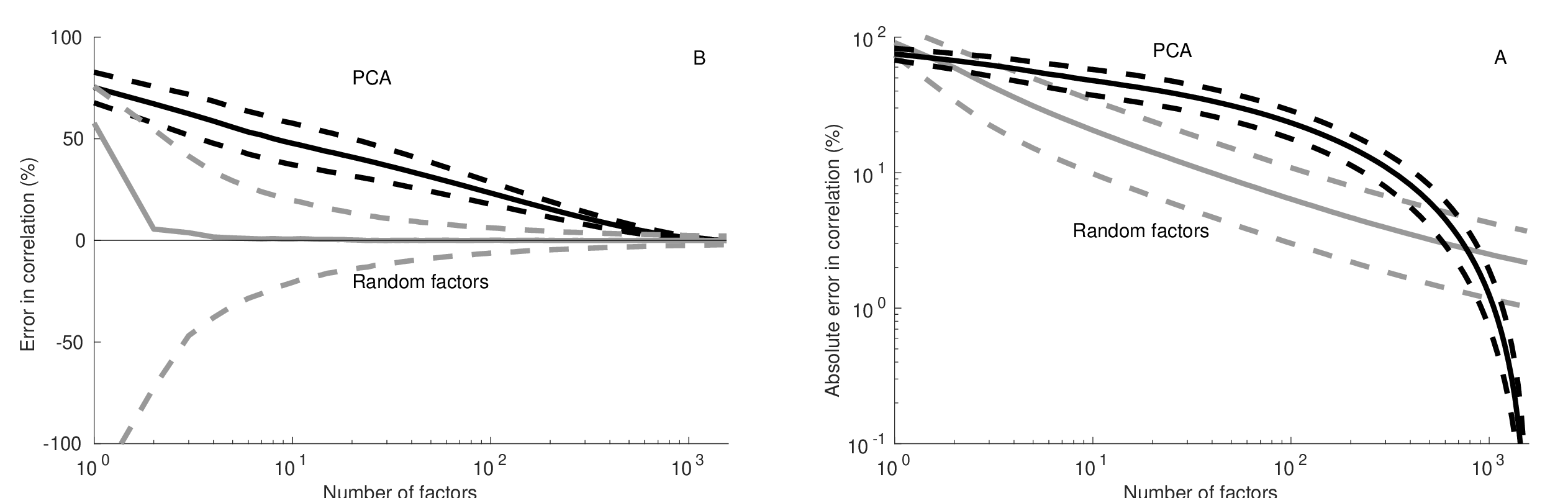}
\caption{Accuracy of the correlation modeling. (A) Median error (solid gray curves; measured in percentage points) 
in correlation coefficient estimates in all pairs in the data-set estimated 
using 1,000 different random factor models, 
together with the 25th and 75th (dashed gray curves) percentiles of error in  correlation estimates. 
The results are compared with the estimates of correlation based on PCA (solid black curve), 
together with the 25th and 75th (dashed black curves) percentiles. 
The results are shown as a function of the number of factors (abscissa).  
(B) Median absolute error of the random factor model (solid grey curve; measured in percentage points),
together with the 25th and 75th percentiles (dashed grey curves); Median error in PCA (solid black curve)
and the 25th and 75th percentiles (dashed black curves).}
\label{fig:fig3}
\end{figure}

\subsection{Correlation coefficient}
 
Fig.\ 3 shows the accuracy of reproduction of correlation coefficients in all analyzed pairs of stocks. 
In the RFM, the median error converges rapidly to zero with only a few factors. 
The 25th and 75th percentiles of error converge toward zero when the number of factors increases.
Together these three curves form a funnel (Fig.\ 3A) that rapidly converges toward zero. 
This shows that the typical accuracy of the correlation coefficient reproduction improves rapidly with the number of factors.
Still, some noise persists even with the ``full'' set of factors, for $k=d$.

In PCA, median error approaches zero only with around 1,000 factors, 
which is largely a consequence of PCA significantly underestimating volatilities of time-series. 
The 25th and 75th percentiles concentrate around the median away from zero in PCA.

Fig.\ 3B shows results on absolute error in correlation coefficient\footnote{Absolute 
error is defined as the absolute difference $|\cmodel-\cdata|$.} as a function of the number of factors. 
In the RFM, correlation estimates converge toward the exact value when the number of factors is increased, 
however, convergence is less rapid than in analysis shown in Fig.\ 3A. 
This is a result of the fact that error can be in either direction in the RFM.
Compared with PCA, correlation estimates in the RFM 
converge significantly more rapidly toward the exact value. 
Since error is always in the same direction in PCA, there are no differences
between absolute error and relative error in PCA-based analyses.

The RFM provides a more accurate description of correlation coefficients than PCA, 
when the number of factors is less than about 500. Noise inherent in the random factor 
model has the consequence that the error in correlation estimates does not disappear in 
the RFM even with the full set of variables even though median 
estimate rapidly converges toward the observed correlation.  

The cross-over to regime where PCA is more accurate occurs around 700--800 factors (Fig. 3). 
When the number of factors is very high, PCA gives as good 
as or better correlation coefficients than the median estimate from the RFM. 
A factor model with this large number of factors is of little use in practical applications.

\begin{figure} 
\includegraphics[width=15cm,scale=0.5]{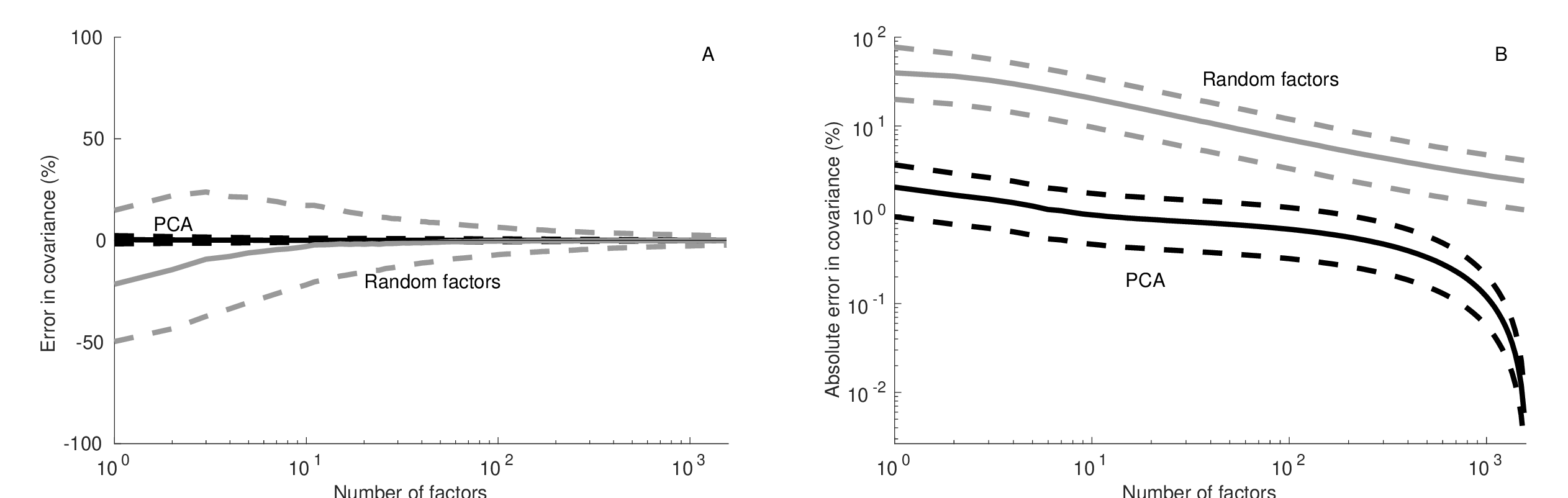}
\caption{Accuracy of covariance estimation. (A) Median error (solid gray curves; measure in percentage points) 
in covariance in all pairs in the data-set estimated using 1,000 different 
random factor models, together with 25th and 75th (dashed gray curves) 
percentiles of error in covariance estimates. The results are compared 
with the estimates of covariance based on PCA (solid black curve), 
together with the 25th and 75th (dashed black curves) percentiles. 
The results are shown as a function of the number of factors (abscissa).  
(B) Median absolute error (solid gray curves; measure in percentage points) 
in covariance in all pairs in the data-set estimated using 1,000 different 
random factor models, together with the 25th and 75th (dashed gray curves) 
percentiles of error. The results are compared 
with the estimates of covariance based on PCA (solid black curve), 
together with the 25th and 75th (dashed black curves) percentiles.}
\label{fig:fig4}
\end{figure}

\subsection{Covariance}

The median error in covariance estimates converges rapidly toward zero in the RFM.
The 25th and 75th percentiles form a funnel that converges toward zero when the number factors increases (Fig.\ 4A).
Despite the fact that PCA is worse than the RFM in reproducing correlation coefficients,
PCA gives a better reproduction of the covariance matrix (Fig.\ 4B).

\begin{figure} 
\includegraphics[width=15cm]{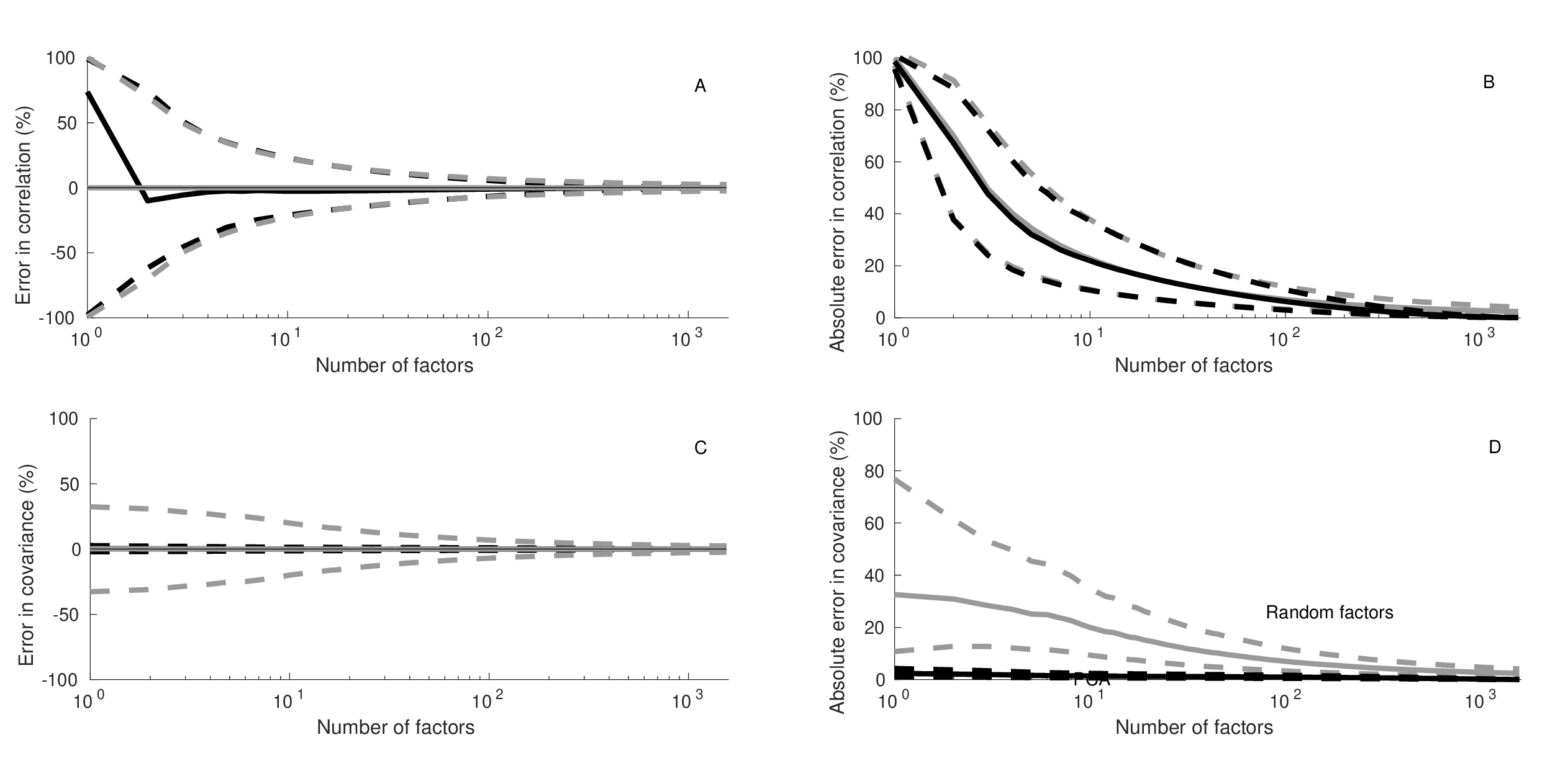}
\caption{Analysis of the reduced data set in which the influence of the market is removed. 
(A) Error in correlation coefficient, (B) Absolute error in correlation coefficient estimates,
(C) Error in covariance, and (D) Absolute error in covariance estimates in an random factor model (gray curves) and in PCA (black curves). 
Solid lines are median estimates, dashed lines 25th and 75th percentiles.}
\label{fig:fig5}
\end{figure}

\begin{figure} 
\includegraphics[width=15cm]{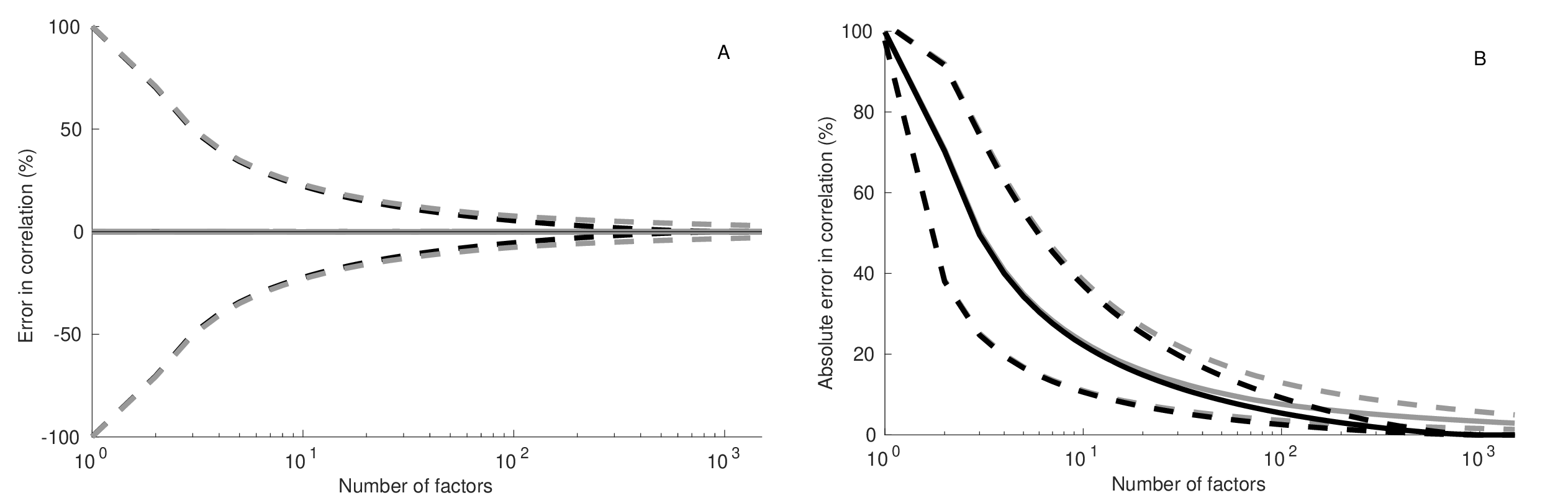}
\caption{Reproduction of correlation coefficient in randomly-generated data. 
(A) Error in correlation reproduction of the random data using the random factor model (dashed gray curve) and PCA (black solid curve) . 
Curves are shown as functions of the number of factors. Differences are in percentage points. 
(B) Absolute error in correlation reproduction of the random data using the random factor model (dashed gray curve) and PCA (black solid curve) . 
}
\label{fig:fig7}
\end{figure}

\subsection{Impact of the market factor}
The risk in the equity market is often dominated by a single factor known as the market risk factor (e.g., Sharpe, 1964).
To better analyze the other possible risk factors, we subtract the first principal component, 
corresponding to the market risk factor, from the data and reanalyze the remaining data (the "reduced data").

Fig.\ 5A shows that PCA becomes more accurate in reproducing the correlation coefficients 
when the impact of the market risk factor is removed from the data. 
Perhaps more surprisingly, reproduction of data structure becomes equally accurate in the RFM and in PCA
with respect to both error measures in the correlation coefficient (Fig.\ 5A and 5B). 
This suggests that the RFM and PCA contain equal amounts of information about the correlations.

As a further check, we generated random data by sampling the normal probability distribution $N(0,1)$ repeatedly. 
Fig.\ 6 shows that the accuracy of both the RFM and PCA is almost identical in this case. 
Comparison with Figures\ 5A and\ 5B shows that
the accuracy of reproduction of the ``reduced'' Russell correlations does not 
significantly differ from the accuracy of the random data. 
This indicates that the fluctuations around the market risk factor are 
largely a product of independent ``noise'' contributions.

Removal of the market risk factor from the data also influences the accuracy of covariance reproduction.
PCA is again more accurate than the RFM in covariance reproduction (Fig.\ 5C and 5D).
In this case, the median error of covariance matrix reproduction does not deviate from zero in the RFM,
and the 25th and 75th percentiles are almost symmetrically around x-axis.

\section{Universality}\label{sec:universality}

A number of probability distributions have been found useful in 
the random projection method (e.g., \cite{achlioptas2003database, kaski1998dimensionality}). 
\cite{matouvsek2008variants} found that almost any probability distribution with zero mean, 
unit variance, and subgaussian tail fulfills the requirements of the Johnson-Lindenstrauss theorem. 
These findings suggest that it may not matter much which probability distribution is used
in the random projections.
To find out whether this is the case in an RFM, we reanalyze the data
using RFMs based on six different probability distributions.  
We have also discussed 
some lowest order effects of varying the probability distribution, as well as reasons
why deviation from a Gaussian distribution leads  only to small corrections, in Remark \ref{th:nonGauss}
after the proof in the Appendix.

\subsection{Probability distributions}
The six probability distributions that we employ here are two sparse matrix models 
of \cite{achlioptas2003database}, a column-normalized Gaussian model, 
a row-normalized Gaussian model, the baseline Gaussian model (defined in Sec. \ref{sec:randomfactormodel}) and 
a uniform model. 
In each case, the probability distribution is symmetric with respect to the origin 
and such that the expectation is zero. Each probability distribution also has a subgaussian tail.
These RFMs differ from the baseline Gaussian RFM only
by the construction of the random projection matrix $B$, and by the normalization.

\subsubsection{Coin-flipping distributions}
The simplest specification for random projection is the "random coin-flipping" algorithm of  \cite{achlioptas2003database}.
It is defined by choosing each element $B_{pq}$ of matrix $B$ independently according to rule: 
set $B_{pq}=+1$ 
with probability 0.5 and set $B_{pq}=-1$ with probability 0.5. 

The second random projection that  \cite{achlioptas2003database} proposes is based on 
a more sparse projection matrix defined by: 
set $B_{pq}=+1$ with probability 1/6, set $B_{pq}=0$ with probability 2/3 
and set $B_{pq}=-1$ with probability 1/6. Again each element is chosen independently of the other elements.
Based on these random projections, we can define two RFMs.

\subsubsection{Gaussian and uniform distributions}
In addition to the baseline Gaussian RFM, 
we analyze two different RFMs based on the normal distribution.
In the first RFM, matrix $B$ is based on the spherical uniform distribution.
The elements of matrix $B$ are defined by 
\begin{equation}
B_{ml}=z_{ml}/Z,
\end{equation}
where $z_{ml}\sim N(0,1)$ are indendent and $Z=\sqrt{\sum\nolimits_p|z_{pl}|^2}$.
In this RFM, columns of matrix $B$ are normalized in such a way that their 
length is exactly one. 

Due to normalization of the columns of matrix $B$, 
diagonal elements of matrix $\BTB\in \Rdd$ behave as in an orthogonal matrix.
Then $(\BTB)_{mm}=1$ for all $m=1,2,\ldots,k$. 
Non-diagonal elements of $\BTB$ have zero expectation and variance 
proportional to $1/d$ (Kaski, 1998). Hence, non-diagonal elements of $\BTB$ are approximately 
distributed according to zero-mean normal distribution at a relatively low dimension. 
Therefore $\BTB=1+\epsilon$, where $\epsilon \in\Rdd$ has non-zero elements only on off-diagonal, 
$\E[\epsilon]=0$ and $|\E[\epsilon]|<2/d$. Matrix $B$ is then almost orthonormal. 

The second RFM based on the Gaussian probability distribution is a variation on the theme:
Instead of column-normalization in the first model, rows of projection matrix $B$ are normalized to unit length.
This is the only difference between the two RFMs, but
it is sufficient to require a different normalization constant.

The sixth considered RFM is defined by projection matrix B, which is based on 
the continuous uniform probability distribution. 
Each element in the projection matrix $B$ is chosen independently
from the uniform distribution on interval $[-1,1]$,
that is, $B_{mn}\sim U(-1,1)$ for each $m,n$.

\begin{figure} 
\includegraphics[width=15cm]{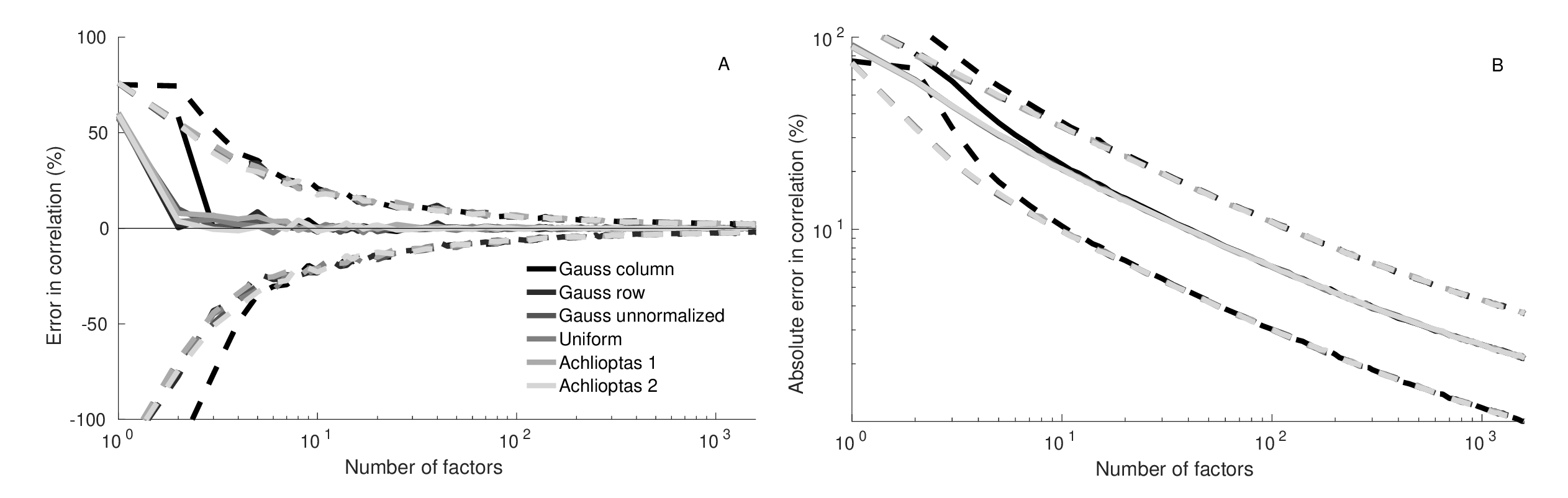}
\caption{Comparison of six different projection matrix specifications. Solid lines are median estimates, dashed lines the 25th
and 75th percentiles. (A) Error in correlation coefficient estimates as functions of the number of factors in six models. Error is computed from the entire set of correlation pairs in 1,591 time-series. 
(B) Absolute error in correlation coefficient estimates in the six models as functions of the number of factors.}
\label{fig:fig6}
\end{figure}

\subsection{Universality of distributions}
Fig.\ 7 shows that all six RFMs produce almost equally accurate results.
To reduce noise, Fig.\ 7 shows results averaged over 50 sample runs.
When the number of factors exceeds 10, all RFMs produce almost identical median accuracy.
The only deviation is the column-normalized Gaussian model, 
which deviates from the other RFMs when the number of factors is less than 5.
All the other RFMs produce identical results also in this regime. 
The accuracy of the 25th and 75th percentiles mainly depends on 
the number of factors, not much on the way factors are generated.

The results suggest that the details of how the projection matrix
is specified are not that important.
Almost any sufficiently regular construction of the random projection matrix (when properly normalized)
produces a factor model, which preserves the approximate correlation structure. 
The main requirement here seems to be that matrix elements
are chosen randomly and independently of other matrix elements.
This supports the view that the RFM represents quite well
how the bulk of factor models would describe the analyzed task.

\section{Discussion}

\subsection{Randomness of factors}

We set out to analyze the impact of random selection of factors on a linear factor model. 
We were interested in whether and how randomness in the choice of factors impacts 
the reproduction of long equity time-series and, in particular, whether their interdependence is preserved. 
We found that accuracy of a typical random factor model is respectable, especially  
the correlation matrix is well-preserved in the reproduction of time-series (Section 4).
We also derived novel theoretical results on the accuracy of a random factor model (Appendix A).

It may seem unlikely that a factor model with randomly-chosen factors could be used for any kind of factor modeling. 
One of the reasons for the ability of an RFM to capture the details of
an equity time-series resides in the fact that random factors are, 
as a consequence of independence of elements, almost orthogonal to each other. 
Furthermore, the number of almost orthogonal vectors is higher in a higher-dimensional space\footnote{This 
finding is often attributed to \cite{hecht1994context}.},
which reduces the impact of the "curse of dimensionality"  \citep{Bellman1957, indyk1998approximate}
and thereby makes data represention more feasible. 
A suitably high number of random factors will then span a  
subspace sufficient to capture the return time-series at the desired accuracy. 

On the other hand, the number of factors is not bounded in a random factor model. 
Only in the limit of infinite number of factors, an RFM is ensured to reproduce the original time-series perfectly.
This can be viewed as a disadvantage of using an RFM.

\subsection{Universality}

In a classical factor model, only a few factors are statistically significant. 
Then, explanatory power of each factor should be large.
In a statistical factor model, a larger number of factors is often used,
which also has the consequence that a larger ambiguity 
in the choice of factors is encountered  \citep{ledoit2004honey}. Several different
sets of factors may provide almost equally good fit to the data.
In an RFM, each factor has only a small explanatory power,
which suggests that a large number of factor sets provide 
essentially equally good descriptions of the data and its structure.
This was observed in our computational experiments.

The number of random factors seems to be more important than
fine-tuning of random factor time-series.
The way an RFM is constructed is not important as long as elements of 
projection $B$ are independently drawn from a suitably regular probability 
distribution with zero expectation and subgaussian tails. 
Regardless of the probability distribution used, we obtained almost 
identically accurate results. These findings suggest that a kind of universality 
of RFMs is present, at least with respect to correlation coefficients.
The results are largely dominated by a set of typical
RFMs that have a rather similar accuracy of data reproduction. 
We have called this set of factor models the bulk.

The analysis of the proof of Theorem A.1 (see Remark A.4) supports 
the view that universality is present with respect to probability distributions.
The assumption that probability distribution is Gaussian is 
not necessarily required in the theorem. It suffices to assume independence of the random matrix elements,
and it is likely that this requirement can be relaxed further.

\subsection{Accuracy, revisited}

In our analyses, we employed PCA as a yardstick against which we compared the RFM. 
The RFM described correlation structures and volatility well, 
but individual data points of time-series were reproduced less accurately. 
PCA reproduced individual data points of time-series more accurately, 
but reproduction of cross-correlations of the time-series was not that good
mainly due to underestimated volatility.
In other words, the RFM preserved the structure of the data 
but not necessarily the details of single time-series, while PCA representation 
preserved the details but not necessarily the correlation structure.

It is worth pointing out that PCA is fitted to preserve co-variance of time-series, not correlations.
An RFM is fitted in no way to the data, so the preservation of
correlations is quite unexpected.

The previous literature has compared the performance of the random projection method with PCA, and found
results similar to ours. 
\cite{bingham2001random} found that random projection method performed significantly better 
than PCA in the compression of image data and in text clustering. 
\cite{goal2005face} found that random projection compares favorably with PCA, 
although PCA is more accurate with small number of dimensions. 
\cite{tang2005comparing} found that in text clustering a PCA-based method provides 
better accuracy with small number of dimensions, while with high number of dimensions the 
random projection method dominated. 
\cite{deegalla2006reducing} found that in five image data sets and five micro array data sets, 
PCA dominated with a small number of dimensions but its performance deteriorated
when the dimensions of the data increased (cross-over occurs at 15--150 dimensions depending on the data set), 
while random projection dominates at high number of dimensions. Our findings are consistent with the results of these previous studies. 

The realm of RFMs is the domain of huge data sets consisting of large number of long time-series.
An RFM answers the question: how many factors will a generic linear factor model
require to describe the data at a specific accuracy.

\section*{Acknowledgements}
The authors thank Dr.\ Petri Niininen for collecting the initial data set.  
The research of J.~Lukkarinen has been supported by the Academy of Finland 
via the Centre of Excellence in Analysis and Dynamics Research (project 271983) and from an Academy Project (project 258302).
This work has also benefited from the support of the project EDNHS ANR-14-CE25-0011 of the French National Research Agency (ANR). 
JL is grateful to Antti Knowles for discussions and references concerning random matrix models.

\appendix

\renewcommand*{\appendixname}{Appendix }
\section{Accuracy of random factor approach}
\renewcommand*{\appendixname}{}

We prove here the following result about the mean and variance of the projection operators involved in the random factor models.
\begin{theorem}\label{th:main}
 Suppose $k\ge 1$, $d \ge 2$ and that the matrix elements of the random matrix $B\in \R^{k\times d}$ 
 are i.i.d.\ and $N(0,1)$-distributed.  For some given $a>0$ define a new random matrix 
 $P\in \R^{d\times d}$ by $P:= a B^T B$.
 
  Then for every non-random vectors $u,v\in \R^d$ all of the following results hold:
 \begin{enumerate}
  \item $\E\!\left[(P u)_m\right] = a k u_m$ for all $m$, and $\E[\mu_{Pu}]= a k \mu_u$.
  \item $\Var((P u)_m) = a^2 k \left(u_m^2 + |u|^2\right)$ for all $m$.
  \item $\E\!\left[C_{Pu,Pv}\right]= a^2 k (d+k) C_{u,v} + a^2 k d \mu_u \mu_v$.
  \item If $\mu_u=\mu_v=0$ and $d\ge 4$, then 
  $\Var\!\left(C_{Pu,Pv}\right)\le a^4 8 k (k+d)^2 \sigma_u^2 \sigma_v^2$.
 \end{enumerate}
\end{theorem}

The main application of the theorem is the following consequence of the above result:
\begin{corollary}\label{th:coroll}
 Suppose $k\ge 1$, $d \ge 4$ and that the matrix elements of the random matrix $B\in \R^{k\times d}$ 
 are i.i.d.\ and $N(0,1)$-distributed.  Set $a=[k (k+d)]^{-1/2}$ and define $P= a B^T B$.

 Then for every $b>0$ and non-random vectors $u,v\in \R^d$, with $\mu_u=\mu_v=0$, all of the following results hold:
 \begin{enumerate}
  \item $\E\!\left[(P u)_m\right] = \left[k/(k+d)\right]^{1/2} u_m$ for all $m$ and $\E[\mu_{Pu}]=0$.
  \item $\Var((P u)_m) = \left(u_m^2 + (d-1) \sigma_u^2\right)/(d+k)$ for all $m$.
  \item $\E\!\left[C_{Pu,Pv}\right]= C_{u,v}$.
  \item $\mathbbm{P}\!\left[|C_{P u,P v}-C_{u,v}|\ge b\right]\le \frac{8}{k b^2} \sigma_u^2 \sigma_v^2$.
 \end{enumerate} 
\end{corollary}
{\it Proof of the Corollary:}\ \ 
 With the added assumptions $\mu_u=\mu_v=0$, the definition of the sample variance yields the identities $|u|^2=(d-1) \sigma_u^2$
and $|v|^2=(d-1) \sigma_v^2$.  Thus items 1, 2 and 3 are immediate corollaries using $a^2 = 1/(k(d+k))$.
On the other hand, if we denote the standard deviation of $C_{Pu,Pv}$ by $S$,
then by Chebyshev's inequality we have 
$\mathbbm{P}\!\left[|C_{Pu,Pv}-C_{u,v}|\ge cS\right]\le c^{-2}$ for any $c>0$.  Applying this 
with $c:=b/S$ thus implies 
$\mathbbm{P}\!\left[|C_{Pu,Pv}-C_{u,v}|\ge b\right]\le S^2 b^{-2}$.  Hence, item 4 of the Theorem
implies the bound in the last item of the Corollary.%\qed

\medskip

\noindent
{\it Proof of the Theorem:}\ \ 
Assume $k,d,B,a,P,u,v$ be given as in the theorem, and define $Z:=P u$, $Z':= Pv$ and $C:=C_{P u,P v}$.
$Z_m$, $Z'_m$, $m=1,2,\ldots,d$, and $C$ are then all real-valued random variables. 

The following results could also be proven by straightforward but rather lengthy direct estimates 
relying on Wick's product rule, i.e., Isserlis' theorem, valid for the present Gaussian
random variables $B_{jm}$.  A better control of the associated
combinatorics is obtained by using Wick polynomial expansions instead.
For the present case of Gaussian random variables Wick polynomials
reduce to Hermite polynomials;
Appendix A of  \citep{lukkarinen2016wick} provides a quick summary of
the definition and main properties of general Wick polynomials, and we refer to  \cite{janson1997gaussian, peccati2011wiener, lukkarinen2016summability}
for more detailed expositions.

We rely on the following Wick polynomial expansion of arbitrary centered expectations of monomials of random variables: for any product of random variables $x_1,x_2,\ldots,x_n$, with the corresponding index set $I=\{1,2,\ldots,n\}$, one has
\begin{align}\label{eq:Wickpexp}
 \prod_{i=1}^n x_i - \E\biggl[\prod_{i=1}^n x_i\biggr] 
 = \sum_{\emptyset\ne E \subset I} \E\biggl[\prod_{i\in I\setminus E} x_i\biggr] \w{\prod_{i\in E} x_i}\, .
\end{align}

For any $m=1,2,\ldots,d$, the definitions yield
\begin{align}
 Z_m = (Pu)_m = a \sum_{n=1}^d \sum_{j=1}^k B_{jm} B_{jn} u_n\, .
\end{align}
Since $B_{jm}$ are i.i.d.\ centered, normalized Gaussian, this implies
\begin{align}
 \E[Z_m] = a \sum_{n=1}^d \sum_{j=1}^k \cf{n=m} u_n = a k u_m\, ,
\end{align}
where $\cf{n=m}$ stands for an indicator function having value $1$, when $n=m$, and $0$ otherwise.
Thus $\E[\mu_{Pu}] = \frac{1}{d} \sum_{n=1}^d \E[Z_m] = a k \mu_u$
As $\E[B_{jm}]=0$, centering the variable $Z_m$ yields the following simple Wick polynomial expansion:
\begin{align}\label{eq:centredZ}
 Z_m - \E[Z_m] = a \sum_{n=1}^d  u_n \sum_{j=1}^k \w{B_{jm} B_{jn}}\, .
\end{align}

The usefulness of Wick polynomials lies in the property that their products satisfy the same moments-to-cumulants expansion as 
simple products, with the additional rule that {\em any partition with a cluster of indices inside one of the Wick polynomials will be missing} from the expansion.  For instance, for any random variables $x_1,x_2,x_3,x_4$---which need not be independent nor Gaussian---one has
\begin{align}
 \E[\w{x_1 x_2}\w{x_3 x_4}] = \kappa[x_1,x_2,x_3,x_4] + \kappa[x_1,x_3]\kappa[x_2,x_4] +
 \kappa[x_1,x_4]\kappa[x_2,x_3] \, ,
\end{align}
where $\kappa$ denotes a cumulant.  Here, for instance, $\kappa[x_1,x_3]=\ccov(x_1,x_3)$.  Applying this to the $B$-variables yields 
\begin{align}\label{eq:fourBs}
& \E[\w{B_{jm} B_{jn}}\w{B_{j'm'} B_{j'n'}}]
= \cf{j'=j,m'=m}\cf{j'=j,n'=n} +
\cf{j'=j,n'=m}\cf{j'=j,n=m'} \, ,
\end{align}
since for Gaussian random variables the fourth cumulant is equal to zero.
Therefore, by (\ref{eq:centredZ}) and (\ref{eq:fourBs}), we have
\begin{align}\label{eq:zzpcov}
& \ccov(Z_m,Z'_{m'}) = \E[(Z_m - \E[Z_m])(Z'_{m'} - \E[Z'_{m'}])] 
\nonumber\\ & \quad
= a^2 \sum_{n',n=1}^d  u_n  v_{n'} \sum_{j',j=1}^k \E[\w{B_{jm} B_{jn}}\w{B_{j'm'} B_{j'n'}}]
% \nonumber\\ & \quad
% =a^2 \sum_{n',n=1}^d  u_n  v_{n'} \sum_{j',j=1}^k \left(\cf{j'=j,m'=m}\cf{j'=j,n'=n} +
% \cf{j'=j,n'=m}\cf{j'=j,n=m'})\right) 
\nonumber\\ & \quad
=a^2 \left(k \cf{m'=m}\sum_{n=1}^d  u_n v_n + k u_{m'} v_{m}\right) =
a^2 k \left(\cf{m'=m} u\cdot v + u_{m'} v_{m}\right)\, ,
\end{align}
and thus, in particular, 
\begin{align}
& \Var(Z_m) = \E[(Z_m - \E[Z_m])^2] 
=  a^2 k \left(|u|^2 +   u_m^2\right)\, .
\end{align}
Therefore, we have now proven the first two items of the Theorem.

The combinatorics gets progressively heavier in the remaining two items.  
Let us begin with the scalar product
\begin{align}
 & Pu\cdot Pv = Z\cdot Z'
 = \sum_{m=1}^d Z_m Z'_m 
 %\nonumber\\ & \quad
 = \sum_{m=1}^d z_m z'_m + 
 a k \sum_{m=1}^d (u_m z'_m+ v_m z_m) + 
  a^2 k^2 \sum_{m=1}^d u_m v_m
 \, ,
\end{align}
where $z_m=Z_m-\E[Z_m]=Z_m-a k u_m$, $z'_m=Z'_m-\E[Z'_m]=Z'_m-a k v_m$ denote the centered variables.  
Taking an expectation and using (\ref{eq:zzpcov})
for $m'=m$ thus yields 
\begin{align}
 & \E[Pu\cdot Pv]= \sum_{m=1}^d 
a^2 k \left(u\cdot v + u_{m} v_{m}\right)
+ a^2 k^2  u\cdot v
= a^2 k (d+ 1 +k) u\cdot v 
 \, .
\end{align}

The definition of $C$ reads explicitly
\begin{align}
 & C = C_{P u,P v} = \frac{1}{d-1} \sum_{m=1}^d (Pu)_m (Pv)_m - \frac{d}{d-1} \mu_{Pu} \mu_{Pv}
 = \frac{1}{d-1} Z \cdot Z' - \frac{d}{d-1} \mu_{Z} \mu_{Z'}
 \, .
\end{align}
To compute its expectation, we still need to evaluate
\begin{align}
& \E[ \mu_{Z} \mu_{Z'} ] = \ccov [\mu_{Z}, \mu_{Z'}] + \E[ \mu_{Z} ] \E[ \mu_{Z'} ] 
% \nonumber\\ & \quad
= \frac{1}{d^2}\sum_{m',m=1}^d \ccov [Z_m, Z'_{m'}] + a^2 k^2 \mu_u \mu_v
 \nonumber\\ & \quad
= \frac{a^2 k }{d^2}\sum_{m',m=1}^d \left(\cf{m'=m} u\cdot v + u_{m'} v_{m}\right) + a^2 k^2 \mu_u \mu_v
% \nonumber\\ & \quad
= \frac{a^2 k }{d} u\cdot v + a^2 k ( k+1) \mu_u \mu_v\, .
\end{align}
Therefore,
\begin{align}
 & \E[C] = \frac{a^2 k}{d-1} (d+k) u\cdot v - \frac{d}{d-1} a^2 k ( k+1) \mu_u \mu_v
 = a^2 k (d+k) C_{u,v} + a^2 k d \mu_u \mu_v
 \, ,
\end{align}
where in the last step we have used the identity $u\cdot v = (d-1) C_{u,v}+ d \mu_u\mu_v$.

For the final result, let us assume in addition that $\mu_u=\mu_v=0$.  To avoid iterated Wick polynomials, let us begin with 
$C=\frac{1}{d-1} Z \cdot Z' - \frac{d}{d-1} \mu_{Z} \mu_{Z'}$ and express the two terms separately in Wick form.
Namely, now
\begin{align}
\mu_{Z} \mu_{Z'} = \frac{a^2}{d^2} \sum_{n',n=1}^d v_{n'} u_n 
\sum_{m',m=1}^d \sum_{j',j=1}^k B_{jm} B_{jn} B_{j'm'} B_{j'n'}\, ,
\end{align}
where the product of four $B$-factors can be expanded using Wick polynomial expansion (\ref{eq:Wickpexp}).  Since only expectations of 
products of even number of $B$:s can be non-zero, we obtain
\begin{align}\label{eq:fourBWick}
 & B_{jm} B_{jn} B_{j'm'} B_{j'n'} - \E[B_{jm} B_{jn} B_{j'm'} B_{j'n'}]
 \nonumber\\ & \quad
 = \w{B_{jm} B_{jn} B_{j'm'} B_{j'n'}} 
 + \E[B_{jm} B_{jn}]\w{B_{j'm'} B_{j'n'}} 
 + \E[B_{jm} B_{j'm'}]\w{ B_{jn} B_{j'n'}} 
 \nonumber\\ & \qquad
 + \E[B_{jm} B_{j'n'} ]\w{B_{jn} B_{j'm'} } 
 + \E[B_{jn} B_{j'm'}]\w{ B_{jm}  B_{j'n'}} 
 \nonumber\\ & \qquad
 + \E[B_{jn} B_{j'n'}]\w{B_{jm}  B_{j'm'} } 
 + \E[B_{j'm'} B_{j'n'}]\w{B_{jm} B_{jn}} 
 \nonumber\\ & \quad
 = \w{B_{jm} B_{jn} B_{j'm'} B_{j'n'}} 
 + \cf{m=n}\w{B_{j'm'} B_{j'n'}} 
 + \cf{m'=n'}\w{B_{jm} B_{jn}} 
 \nonumber\\ & \qquad
 + \cf{j'=j}\bigl[
 \cf{m=m'}\w{ B_{jn} B_{j n'}} 
 + \cf{m=n'} \w{B_{jn} B_{j m'} } 
 %~ \nonumber & \qquad\quad
 + \cf{m'=n} \w{ B_{jm}  B_{j n'}} 
 + \cf{n=n'} \w{B_{jm}  B_{jm'} } \bigr]
% \nonumber\\ & \qquad
\end{align}
Therefore,
\begin{align}
& \mu_{Z} \mu_{Z'} - \E[\mu_{Z} \mu_{Z'}]
% = \frac{a^2}{d^2} \sum_{n',n=1}^d v_{n'} u_n \sum_{m',m=1}^d \sum_{j',j=1}^k \w{B_{jm} B_{jn} B_{j'm'} B_{j'n'}} 
%  \nonumber\\ & \qquad
% + \frac{a^2}{d^2} \sum_{n',n=1}^d v_{n'} u_n \sum_{m'=1}^d \sum_{j'=1}^k k \w{B_{j'm'} B_{j'n'}} 
% % \nonumber\\ & \qquad
% + \frac{a^2}{d^2} \sum_{n',n=1}^d v_{n'} u_n \sum_{m=1}^d \sum_{j=1}^k k \w{B_{jm} B_{jn}} 
%  \nonumber\\ & \qquad
%  + \frac{a^2}{d^2} \sum_{n',n=1}^d v_{n'} u_n \sum_{j=1}^k d \w{ B_{jn} B_{j n'}} 
% % \nonumber\\ & \qquad
%  + \frac{a^2}{d^2} \sum_{n',n=1}^d v_{n'} u_n \sum_{m'=1}^d \sum_{j=1}^k \w{B_{jn} B_{j m'} } 
%  \nonumber\\ & \qquad
%  + \frac{a^2}{d^2} \sum_{n',n=1}^d v_{n'} u_n \sum_{m=1}^d \sum_{j=1}^k \w{ B_{jm}  B_{j n'}} 
% % \nonumber\\ & \qquad
%  + \frac{a^2}{d^2} \sum_{n=1}^d v_{n} u_n \sum_{m',m=1}^d \sum_{j=1}^k \w{B_{jm}  B_{j m'} }
% \nonumber\\ & \quad
= \frac{a^2}{d^2} \sum_{n',n=1}^d v_{n'} u_n \sum_{m',m=1}^d \sum_{j',j=1}^k \w{B_{jm} B_{jn} B_{j'm'} B_{j'n'}} 
 \nonumber\\ & \qquad
 + \frac{a^2}{d} \sum_{n',n=1}^d v_{n'} u_n \sum_{j=1}^k \w{ B_{jn} B_{j n'}} 
 + \frac{a^2}{d} \frac{d-1}{d} C_{u,v} \sum_{m',m=1}^d \sum_{j=1}^k \w{B_{jm}  B_{j m'} }\, ,
\end{align}
where we have applied the assumptions $\mu_u=0=\mu_v$.

Similarly, since
\begin{align}
\frac{1}{d-1} Z \cdot Z' = \frac{a^2}{d-1} \sum_{n',n=1}^d v_{n'} u_n 
\sum_{m=1}^d \sum_{j',j=1}^k B_{jm} B_{jn} B_{j'm} B_{j'n'}\, ,
\end{align}
and by (\ref{eq:fourBWick})
\begin{align}
 & B_{jm} B_{jn} B_{j'm} B_{j'n'} - \E[B_{jm} B_{jn} B_{j'm} B_{j'n'}]
 \nonumber\\ & \quad
 = \w{B_{jm} B_{jn} B_{j'm} B_{j'n'}} 
 + \cf{m=n}\w{B_{j'm} B_{j'n'}} 
 + \cf{m=n'}\w{B_{jm} B_{jn}} 
 \nonumber\\ & \qquad
 + \cf{j'=j}\bigl[
 \w{ B_{jn} B_{j n'}} 
 + \cf{m=n'} \w{B_{jn} B_{j m} } 
% \nonumber\\ & \qquad\quad
 + \cf{m=n} \w{ B_{jm}  B_{j n'}} 
 + \cf{n=n'} \w{B_{jm}  B_{j m} } \bigr] \, ,
% \nonumber\\ & \qquad
\end{align}
we obtain a Wick polynomial expansion
\begin{align}
& \frac{1}{d-1} Z \cdot Z' - \E\left[ \frac{1}{d-1} Z \cdot Z'\right]
% = \frac{a^2}{d} \sum_{n',n=1}^d v_{n'} u_n 
% \sum_{m=1}^d \sum_{j',j=1}^k \w{B_{jm} B_{jn} B_{j'm} B_{j'n'}}
%  \nonumber\\ & \qquad
% + \frac{a^2}{d} \sum_{n',n=1}^d v_{n'} u_n 
% \sum_{j'=1}^k k \w{B_{j'n} B_{j'n'}} 
% + \frac{a^2}{d} \sum_{n',n=1}^d v_{n'} u_n \sum_{j=1}^k k \w{B_{jn'} B_{jn}} 
%  \nonumber\\ & \qquad
%  + \frac{a^2}{d} \sum_{n',n=1}^d v_{n'} u_n 
% \sum_{m=1}^d \sum_{j=1}^k
%  \w{ B_{jn} B_{j n'}} 
%  + \frac{a^2}{d} \sum_{n',n=1}^d v_{n'} u_n \sum_{j=1}^k \w{B_{jn} B_{j n'} } 
%  \nonumber\\ & \qquad
%  + \frac{a^2}{d} \sum_{n',n=1}^d v_{n'} u_n  \sum_{j=1}^k \w{ B_{jn}  B_{j n'}} 
%  + \frac{a^2}{d} \sum_{n=1}^d v_{n} u_n 
% \sum_{m=1}^d \sum_{j=1}^k \w{B_{jm}  B_{j m} }
% \nonumber\\ & \quad
 = \frac{a^2}{d-1} \sum_{n',n=1}^d v_{n'} u_n 
\sum_{m=1}^d \sum_{j',j=1}^k \w{B_{jm} B_{jn} B_{j'm} B_{j'n'}}
 \nonumber\\ & \qquad
+ \frac{a^2}{d-1} (2 k +d +2)  \sum_{n',n=1}^d v_{n'} u_n \sum_{j=1}^k \w{B_{jn} B_{jn'}} 
+ a^2 C_{u,v} \sum_{m=1}^d \sum_{j=1}^k \w{B_{jm}  B_{j m} }
\, .
\end{align}

Combining the above results together finally yields a Wick polynomial expansion for 
the centered $C$,
\begin{align}
& C-\E[C]
=  \frac{a^2}{d-1} \sum_{n',n=1}^d v_{n'} u_n 
\sum_{m=1}^d \sum_{j',j=1}^k \w{B_{jm} B_{jn} B_{j'm} B_{j'n'}}
 \nonumber\\ & \qquad
-\frac{a^2}{d (d-1)} \sum_{n',n=1}^d v_{n'} u_n \sum_{m',m=1}^d \sum_{j',j=1}^k \w{B_{jm} B_{jn} B_{j'm'} B_{j'n'}} 
 \nonumber\\ & \qquad
+ \frac{a^2}{d-1} (2 k +d +1)  \sum_{n',n=1}^d v_{n'} u_n \sum_{j=1}^k \w{B_{jn} B_{jn'}} 
 \nonumber\\ & \qquad
+ a^2 C_{u,v} \sum_{m=1}^d \sum_{j=1}^k \w{B_{jm}  B_{j m} }
 - \frac{a^2}{d} C_{u,v} \sum_{m',m=1}^d \sum_{j=1}^k \w{B_{jm}  B_{j m'} }\, .
\end{align}
We use this formula to compute $\Var(C)=\E[(C-\E[C])^2]$.  In the expanded formula terms containing a product of different
degree Wick polynomials yield zero since whatever three pairings is used for the six $B$-factors, one of these pairings connects two 
elements inside the degree four Wick polynomial.  Hence, for instance, 
$\E[\w{B_{j_1m_1} B_{j_1n_1} B_{j'_1m'_1} B_{j'_1n'_1}} \w{B_{j_2m_2}  B_{j_2 m'_2} }]=0$.

The products of second order terms turn out to yield the dominant contribution.  After first taking out a factor $a^4d^{-2} (d-1)^{-2}$, 
it reads explicitly
 %~ \nonumber\\ & \qquad
\begin{align}\label{eq:W2contrib}
& \E\biggl[\biggl( d (2 k +d +1)  \sum_{n',n=1}^d v_{n'} u_n \sum_{j=1}^k \w{B_{jn} B_{jn'}} 
+ d u\cdot v\sum_{m=1}^d \sum_{j=1}^k \w{B_{jm}  B_{j m} }
 - u\cdot v\sum_{m',m=1}^d \sum_{j=1}^k \w{B_{jm}  B_{j m'} }
\biggr)^2 \biggr] 
 \nonumber\\ & \quad
= d^2 (2 k +d +1)^2 \sum_{n'_1,n_1,n'_2,n_2=1}^d v_{n'_1} u_{n_1} v_{n'_2} u_{n_2}  \sum_{j_1,j_2=1}^k
\E[\w{B_{j_1n_1} B_{j_1n'_1}} \w{B_{j_2n_2} B_{j_2n'_2}} ]
 \nonumber\\ & \qquad
+ d^2 (u\cdot v)^2 \sum_{m_1,m_2=1}^d \sum_{j_1,j_2=1}^k \E[\w{B_{j_1 m_1} B_{j_1 m_1}} \w{B_{j_2 m_2} B_{j_2 m_2}} ]
 \nonumber\\ & \qquad
+ (u\cdot v)^2 \sum_{m'_1,m_1,m'_2,m_2=1}^d \sum_{j_1,j_2=1}^k \E[\w{B_{j_1 m_1} B_{j_1 m'_1}} \w{B_{j_2 m_2} B_{j_2 m'_2}} ]
 \nonumber\\ & \qquad
+ 2 d^2 (2 k +d +1) u\cdot v \sum_{n',n=1}^d v_{n'} u_n \sum_{j,j_2=1}^k \sum_{m_2=1}^d \E[\w{B_{jn} B_{jn'}} \w{B_{j_2m_2}  B_{j_2 m_2} }]
 \nonumber\\ & \qquad
- 2 d (2 k +d +1) u\cdot v \sum_{n',n=1}^d v_{n'} u_n \sum_{j,j_2=1}^k \sum_{m_2.m'_2=1}^d \E[\w{B_{jn} B_{jn'}} \w{B_{j_2m_2}  B_{j_2 m'_2} }]
 \nonumber\\ & \qquad
- 2 d (u\cdot v)^2 \sum_{j,j_2=1}^k \sum_{m,m_2.m'_2=1}^d\E[\w{B_{jm}  B_{j m}} \w{B_{j_2m_2}  B_{j_2 m'_2} }],
 %~ \nonumber\\ & \quad
\end{align}
which simplifies to
\begin{align}\label{eq:W2contrib2}
& d^2 (2 k +d +1)^2 k (|u|^2 |v|^2 + (u\cdot v)^2)
+ 2 d^2 (u\cdot v)^2 d k + 2  (u\cdot v)^2  k d^2
 \nonumber\\ & \qquad
 + 4 d^2 k (2 k +d+1)  (u\cdot v)^2 -0 
 - 4 d^2 k (u\cdot v)^2
 \nonumber\\ & \quad
= d^2 (2 k +d +1)^2 k |u|^2 |v|^2 + 
[d^2 (2 k +d +1)^2 k+ 2 d^3 k +2 d^2k + 4 d^2 k (2k+d)]
(u\cdot v)^2
\, .
\end{align}

In the remaining products, the allowed pairings are in one-to-one correspondence to permutations where each factor in the left product is paired with the 
factor in its ``permuted'' position in the right product.  Thus for order four terms we obtain a sum over $4!=24$ terms, namely,
\begin{align}
& \E[\w{B_{j_1m_1} B_{j_1n_1} B_{j'_1m'_1} B_{j'_1n'_1}} \w{B_{j_2m_2} B_{j_2n_2} B_{j'_2m'_2} B_{j'_2n'_2}}] 
 \nonumber\\ & \quad
 = \cf{j_1=j_2,j'_1=j'_2} \bigl[\cf{m_1=m_2,n_1=n_2} + \cf{m_1=n_2,n_1=m_2} \bigr]
 \times
 \bigl[\cf{m'_1=m'_2,n'_1=n'_2} + \cf{m'_1=n'_2,n'_1=m'_2}\bigr]
 \nonumber\\ & \qquad
 + \cf{j_1=j_2=j'_1=j'_2} \cf{m_1=m_2,n_1=m'_2}
% \nonumber\\ & \qquad\quad \times
 \bigl[\cf{m'_1=n_2,n'_1=n'_2} + \cf{m'_1=n'_2,n'_1=n_2}\bigr]
 \nonumber\\ & \qquad
 + \cf{j_1=j_2=j'_1=j'_2} \cf{m_1=m_2,n_1=n'_2}
% \nonumber\\ & \qquad\quad \times
 \bigl[\cf{m'_1=n_2,n'_1=m'_2} + \cf{m'_1=m'_2,n'_1=n_2}\bigr]
 \nonumber\\ & \qquad
 + \cf{j_1=j_2=j'_1=j'_2} \cf{m_1=n_2,n_1=m'_2}
% \nonumber\\ & \qquad\quad \times
 \bigl[\cf{m'_1=m_2,n'_1=n'_2} + \cf{m'_1=n'_2,n'_1=m_2}\bigr]
 \nonumber\\ & \qquad
 + \cf{j_1=j_2=j'_1=j'_2} \cf{m_1=n_2,n_1=n'_2}
% \nonumber\\ & \qquad\quad \times
 \bigl[\cf{m'_1=m_2,n'_1=m'_2} + \cf{m'_1=m'_2,n'_1=m_2}\bigr]
 \nonumber\\ & \qquad
 + \cf{j_1=j'_2,j'_1=j_2} \bigl[\cf{m_1=m'_2,n_1=n'_2} + \cf{m_1=n'_2,n_1=m'_2} \bigr]
 \times
 \bigl[\cf{m'_1=m_2,n'_1=n_2} + \cf{m'_1=n_2,n'_1=m_2}\bigr]
 \nonumber\\ & \qquad
 + \cf{j_1=j_2=j'_1=j'_2} \cf{m_1=m'_2,n_1=m_2}
% \nonumber\\ & \qquad\quad \times
 \bigl[\cf{m'_1=n_2,n'_1=n'_2} + \cf{m'_1=n'_2,n'_1=n_2}\bigr]
 \nonumber\\ & \qquad
 + \cf{j_1=j_2=j'_1=j'_2} \cf{m_1=m'_2,n_1=n_2}
% \nonumber\\ & \qquad\quad \times
 \bigl[\cf{m'_1=n'_2,n'_1=m_2} + \cf{m'_1=m_2,n'_1=n'_2}\bigr]
 \nonumber\\ & \qquad
 + \cf{j_1=j_2=j'_1=j'_2} \cf{m_1=n'_2,n_1=m_2}
% \nonumber\\ & \qquad\quad \times
 \bigl[\cf{m'_1=m'_2,n'_1=n_2} + \cf{m'_1=n_2,n'_1=m'_2}\bigr]
 \nonumber\\ & \qquad
 + \cf{j_1=j_2=j'_1=j'_2} \cf{m_1=n'_2,n_1=n_2}
% \nonumber\\ & \qquad\quad \times
 \bigl[\cf{m'_1=m_2,n'_1=m'_2} + \cf{m'_1=m'_2,n'_1=m_2}\bigr] \, .
\end{align}
Therefore,
\begin{align}
& \sum_{j'_1,j_1,j'_2,j_2=1}^k
\E\biggl[ \w{B_{j_1m_1} B_{j_1n_1} B_{j'_1m'_1} B_{j'_1n'_1}} \w{B_{j_2m_2} B_{j_2n_2} B_{j'_2m'_2} B_{j'_2n'_2}}\biggr]
 \nonumber\\ & \quad
 = k^2 \bigl\{ \bigl[\cf{m_1=m_2,n_1=n_2} + \cf{m_1=n_2,n_1=m_2} \bigr]
 \times
 \bigl[\cf{m'_1=m'_2,n'_1=n'_2} + \cf{m'_1=n'_2,n'_1=m'_2}\bigr]
 \nonumber\\ & \qquad\quad
 +  \bigl[\cf{m_1=m'_2,n_1=n'_2} + \cf{m_1=n'_2,n_1=m'_2} \bigr]
 \times
 \bigl[\cf{m'_1=m_2,n'_1=n_2} + \cf{m'_1=n_2,n'_1=m_2}\bigr] \bigr\}
 \nonumber\\ & \qquad 
 + k \bigl\{ \cf{m_1=m_2,n_1=m'_2}
% \nonumber\\ & \qquad\quad \times
 \bigl[\cf{m'_1=n_2,n'_1=n'_2} + \cf{m'_1=n'_2,n'_1=n_2}\bigr]
 \nonumber\\ & \qquad\quad
 +  \cf{m_1=m_2,n_1=n'_2}
% \nonumber\\ & \qquad\quad \times
 \bigl[\cf{m'_1=n_2,n'_1=m'_2} + \cf{m'_1=m'_2,n'_1=n_2}\bigr]
 \nonumber\\ & \qquad\quad
 +  \cf{m_1=n_2,n_1=m'_2}
% \nonumber\\ & \qquad\quad \times
 \bigl[\cf{m'_1=m_2,n'_1=n'_2} + \cf{m'_1=n'_2,n'_1=m_2}\bigr]
 \nonumber\\ & \qquad\quad
 +  \cf{m_1=n_2,n_1=n'_2}
% \nonumber\\ & \qquad\quad \times
 \bigl[\cf{m'_1=m_2,n'_1=m'_2} + \cf{m'_1=m'_2,n'_1=m_2}\bigr] 
 \nonumber\\ & \qquad\quad
 +  \cf{m_1=m'_2,n_1=m_2}
% \nonumber\\ & \qquad\quad \times
 \bigl[\cf{m'_1=n_2,n'_1=n'_2} + \cf{m'_1=n'_2,n'_1=n_2}\bigr]
 \nonumber\\ & \qquad\quad
 +  \cf{m_1=m'_2,n_1=n_2}
% \nonumber\\ & \qquad\quad \times
 \bigl[\cf{m'_1=n'_2,n'_1=m_2} + \cf{m'_1=m_2,n'_1=n'_2}\bigr]
 \nonumber\\ & \qquad\quad
 +  \cf{m_1=n'_2,n_1=m_2}
% \nonumber\\ & \qquad\quad \times
 \bigl[\cf{m'_1=m'_2,n'_1=n_2} + \cf{m'_1=n_2,n'_1=m'_2}\bigr]
 \nonumber\\ & \qquad\quad
 +  \cf{m_1=n'_2,n_1=n_2}
% \nonumber\\ & \qquad\quad \times
 \bigl[\cf{m'_1=m_2,n'_1=m'_2} + \cf{m'_1=m'_2,n'_1=m_2}\bigr]\bigr\} 
\, .
\end{align}

For the terms involving $Z'\cdot Z$ we also need restrictions of this result to cases where $m_1=m'_1$ or $m_2=m'_2$:
\begin{align}
& \sum_{m_1=1}^d \sum_{j'_1,j_1,j'_2,j_2=1}^k
\E\biggl[ \w{B_{j_1m_1} B_{j_1n_1} B_{j'_1m_1} B_{j'_1n'_1}} \w{B_{j_2m_2} B_{j_2n_2} B_{j'_2m'_2} B_{j'_2n'_2}}\biggr]
 \nonumber\\ & \quad
 = (k^2+k) \bigl[\cf{m'_2=m_2,n_2=n_1,n'_2=n'_1} + 
 \cf{m'_2=n'_1,m_2=n'_2,n_2=n_1} +
 \cf{m'_2=n_2,m_2=n_1,n'_2=n'_1} +
 \nonumber\\ & \qquad\quad
 + 
 \cf{m'_2=n'_1,m_2=n_1,n'_2=n_2} +
 \cf{m'_2=m_2,n_2=n'_1,n'_2=n_1} +
 \cf{m'_2=n_2,m_2=n'_1,n'_2=n_1} 
 \nonumber\\ & \qquad\quad
 +
 \cf{m'_2=n_1,m_2=n'_2,n_2=n'_1} +
 \cf{m'_2=n_1,m_2=n'_1,n'_2=n_2} 
 \bigr]
 \nonumber\\ & \qquad 
 + 2 k \bigl[
  \cf{m'_2=n_1,m_2=n_2,n'_2=n'_1} +
  \cf{m'_2=n'_1,m_2=n_2,n'_2=n_1} 
 \nonumber\\ & \qquad\quad
  + \cf{m'_2=n'_2,m_2=n_1,n_2=n'_1} +
  \cf{m'_2=n'_2,m_2=n'_1,n_2=n_1} 
 \bigr]
\, .
\end{align}
Then we can collect the three terms needed here which are 
\begin{align}
& \sum_{m_1,m_2=1}^d \sum_{j'_1,j_1,j'_2,j_2=1}^k
\E\biggl[ \w{B_{j_1m_1} B_{j_1n_1} B_{j'_1m_1} B_{j'_1n'_1}} \w{B_{j_2m_2} B_{j_2n_2} B_{j'_2m_2} B_{j'_2n'_2}}\biggr]
%  \nonumber\\ & \quad
%  = k(k+1)\sum_{m_2=1}^d \bigl[\cf{n_2=n_1,n'_2=n'_1) + 
%  \cf{n'_1=m_2=n'_2,n_2=n_1) 
%  \nonumber\\ & \qquad\quad
%  + 
%  \cf{m_2=n_2=n_1,n'_2=n'_1) +
%  \cf{n'_1=m_2=n_1,n'_2=n_2) 
%  \nonumber\\ & \qquad\quad
%  +
%  \cf{n_2=n'_1,n'_2=n_1) +
%  \cf{n_2=m_2=n'_1,n'_2=n_1) 
%  \nonumber\\ & \qquad\quad
%  +
%  \cf{n_1=m_2=n'_2,n_2=n'_1) +
%  \cf{n_1=m_2=n'_1,n'_2=n_2) 
%  \bigr]
%  \nonumber\\ & \qquad 
%  + 2 k\sum_{m_2=1}^d \bigl[
%   \cf{n_1=m_2=n_2,n'_2=n'_1) +
%   \cf{n'_1=m_2=n_2,n'_2=n_1) 
%  \nonumber\\ & \qquad\quad
%   + \cf{n'_2=m_2=n_1,n_2=n'_1) +
%   \cf{n'_2=m_2=n'_1,n_2=n_1) 
%  \bigr]
 \nonumber\\ & \quad
 = k(k+1)\bigl[d \cf{n_2=n_1,n'_2=n'_1} + 
 \cf{n'_1=n'_2,n_2=n_1} 
 %\nonumber\\ & \qquad\quad
 + 
 \cf{n_2=n_1,n'_2=n'_1}
 %~ \nonumber\\ & \qquad\quad
 +
 \cf{n'_1=n_1,n'_2=n_2} 
\nonumber\\ & \qquad\quad
 +
 d \cf{n_2=n'_1,n'_2=n_1} +
 \cf{n_2=n'_1,n'_2=n_1} 
 +
 \cf{n_1=n'_2,n_2=n'_1} +
 \cf{n_1=n'_1,n'_2=n_2} 
 \bigr]
 \nonumber\\ & \qquad 
 + 2 k \bigl[
  \cf{n_1=n_2,n'_2=n'_1} +
  \cf{n'_1=n_2,n'_2=n_1} 
  + \cf{n'_2=n_1,n_2=n'_1} +
  \cf{n'_2=n'_1,n_2=n_1} 
 \bigr]
 \nonumber\\ & \quad 
 = [(d+2)k(k+1)+4 k] \cf{n_2=n_1,n'_2=n'_1} 
 + 2k(k+1) \cf{n'_2=n_2,n'_1=n_1} 
 \nonumber\\ & \qquad 
+ [(d+2)k(k+1)+4 k] \cf{n_2=n'_1,n'_2=n_1} 
\, ,
\end{align}
\begin{align}
& \sum_{m_1,m_2,m'_2=1}^d \sum_{j'_1,j_1,j'_2,j_2=1}^k
\E\biggl[ \w{B_{j_1m_1} B_{j_1n_1} B_{j'_1m_1} B_{j'_1n'_1}} \w{B_{j_2m_2} B_{j_2n_2} B_{j'_2m'_2} B_{j'_2n'_2}}\biggr]
 \nonumber\\ & \quad
 = k(k+1) \bigl[d \cf{n_2=n_1,n'_2=n'_1} + 
 \cf{n_2=n_1} 
% \nonumber\\ & \qquad\quad
 + 
 \cf{n'_2=n'_1} +
 \cf{n'_2=n_2} 
 \nonumber\\ & \qquad\quad
 + d
 \cf{n_2=n'_1,n'_2=n_1} +
 \cf{n'_2=n_1} 
% \nonumber\\ & \qquad\quad
 +
 \cf{n_2=n'_1} +
 \cf{n'_2=n_2} 
 \bigr]
 \nonumber\\ & \qquad 
 + 2 k \bigl[
  \cf{n'_2=n'_1} +
  \cf{n'_2=n_1} 
% \nonumber\\ & \qquad\quad
  + \cf{n_2=n'_1} +
  \cf{n_2=n_1}
 \bigr]
\, ,
\end{align}
and
\begin{align}
& \sum_{m_1,m'_1,m_2,m'_2=1}^d\sum_{j'_1,j_1,j'_2,j_2=1}^k
\E\biggl[ \w{B_{j_1m_1} B_{j_1n_1} B_{j'_1m'_1} B_{j'_1n'_1}} \w{B_{j_2m_2} B_{j_2n_2} B_{j'_2m'_2} B_{j'_2n'_2}}\biggr]
 \nonumber\\ & \quad
 = k^2 \bigl[ d^2 \cf{n_1=n_2,n'_1=n'_2} + d \cf{n_1=n_2} + d \cf{n'_1=n'_2} +1
 %~ \nonumber\\ & \qquad\quad
 + d^2 \cf{n_1=n'_2,n'_1=n_2} + d \cf{n_1=n'_2} + d \cf{n'_1=n_2} +1\bigr]
 \nonumber\\ & \qquad 
 + k 
 \bigl[3 d \cf{n'_1=n'_2} + 3 d \cf{n'_1=n_2} + 3 d \cf{n_1=n'_2} +
 3 d \cf{n_1=n_2}+
 \nonumber\\ & \qquad   \quad
 +2 +  d^2 \cf{n_1=n'_2,n'_1=n_2}
+   d^2 \cf{n'_1=n'_2,n_1=n_2}\bigr]
\, .
\end{align}

This yields
\begin{align}
& \E\biggl[ \biggl( d\sum_{n',n=1}^d v_{n'} u_n 
\sum_{m=1}^d \sum_{j',j=1}^k \w{B_{jm} B_{jn} B_{j'm} B_{j'n'}}
 %~ \nonumber\\ & \qquad
- \sum_{n',n=1}^d v_{n'} u_n \sum_{m',m=1}^d \sum_{j',j=1}^k \w{B_{jm} B_{jn} B_{j'm'} B_{j'n'}} 
\biggr)^2 \biggr]
 \nonumber\\ & \quad
= d^2 k \left\{[(d+2)(k+1)+4] |u|^2 |v|^2 + (u\cdot v)^2 [ (d+2)(k+1)+4 + 2(k+1)] \right\}
 \nonumber\\ & \qquad 
+ d^2 k (k+1)(|u|^2 |v|^2 + (u\cdot v)^2)
-2d^2 k (k+1) (|u|^2 |v|^2 + (u\cdot v)^2)
 \nonumber\\ & \quad
= d^2 k [(dk +d+k+5)|u|^2 |v|^2 + (dk+d+3k+7)
(u\cdot v)^2 ]
\, .
\end{align}
Adding the term to (\ref{eq:W2contrib2}), and multiplying the result by $a^4d^{-2} (d-1)^{-2}$ yields
\begin{align}\label{eq:varCresult}
& \Var(C) = \E[(C-\E[C])^2]
 = \frac{a^4}{d^2 (d-1)^2} d^2 k \Bigl[
 (2 k +d +1)^2 |u|^2 |v|^2 
 +(dk +d+k+5)|u|^2 |v|^2 \nonumber\\ & \qquad
+ [ (2 k +d +1)^2 + 2 d +2 + 4 (2k+d)]
(u\cdot v)^2
 + (dk+d+3k+7)
(u\cdot v)^2
 \Bigr]
 \nonumber\\ & \quad
 = \frac{a^4}{(d-1)^2} k \Bigl[
 (4 k^2+d^2+5 dk+5k+3d+6) |u|^2 |v|^2 
 +(4 k^2+d^2+5 dk+15k+9d+10)
(u\cdot v)^2
 \Bigr]
 \nonumber\\ & \quad
 \le\frac{a^4}{(d-1)^2} 2 k  |u|^2 |v|^2 (4 k^2+d^2+5 dk+10 k+6 d+8)
 \nonumber\\ & \quad
 = a^4 2 k \sigma_u^2 \sigma_v^2 (4 (k+d)^2 - 3 d^2
 -3 dk+10 k+6 d+8) \, .
\end{align}

If $d\ge 4$, we have $- 3 d^2-3 dk+10 k+6 d+8\le 0$ for all $k\ge 1$.  Hence, for $d\ge 4$, the above bound can be simplified to the form given in the Theorem,
namely then 
\begin{align}
& \Var(C)\le a^4 8 k (k+d)^2 \sigma_u^2 \sigma_v^2 \, .
\end{align}
This concludes the proof of the theorem.%\qed

\begin{remark}\label{th:constantdisc}
The exact bound in (\ref{eq:varCresult}) can also be approximated in other ways.  
Choosing the normalization as in the Corollary, with $a^2=1/(k(d+k))$, and assuming $d\gg k$,
we obtain an estimate $\Var(C)\lesssim 2 \sigma_u^2 \sigma_v^2/k$ with a reduction of the 
prefactor from $8$ to $2$.  The bound stated in the Theorem becomes optimal in the opposite regime, when $k \gg d$. 
\end{remark}

\begin{remark}\label{th:nonGauss}
The assumption about sufficiently fast decay of correlations, here taken to be i.i.d., between the matrix elements is important for 
the above phenomena to occur.
However, the precise statistics of the distribution of each matrix element plays much less a role.  
For instance, consider instead of Gaussian $N(0,1)$-distributed matrix elements taking them from some other distribution which has finite moments up to order four.  For example, suppose that
the distribution of each $B=B_{jm}$ has mean zero, $\E[B]=0$, a variance $c_2=\E[B^2]$ and a fourth cumulant $c_4=\kappa[B,B,B,B]$.  
As shown below, the resulting changes to the first three items in the Theorem 
are then an introduction of an overall scale $c_2$ and relatively weak 
dependence on $b_4 := c_4/c_2^2$, the excess kurtosis of the distribution of $B$.

Explicitly, instead of equation (\ref{eq:fourBs}) we then have 
\begin{align}%\label{eq:fourBs}
& \E[\w{B_{jm} B_{jn}}\w{B_{j'm'} B_{j'n'}}]
\nonumber\\ & \quad
= c_2^2\bigl( \cf{j'=j,m'=m}\cf{j'=j,n'=n} +
\cf{j'=j,n'=m}\cf{j'=j,n=m'}
%~ \nonumber\\ & \qquad
+ b_4 \cf{j'=j,m'=m=n=n'}\bigr) \, .
\end{align}
Therefore, we obtain
\begin{align}%\label{eq:zzpcov}
& \E[(Pu)_m] = c_2 a k  u_m\, , \\
& \ccov((Pu)_m,(Pv)_{m'}) =
c_2^2 a^2 k \left(\cf{m'=m} u\cdot v + u_{m'} v_{m}+b_4 \cf{m'=m} u_{m} v_{m} \right)\, ,
\end{align}
and, retracing the necessary steps of the above proof thus yields
\begin{align}
& \Var((Pu)_m) = c^2_2 a^2 k \left(|u|^2 + (1+b_4) u_m^2\right)\,, \\
 & c_2^{-2}\E[C_{Pu,Pv}] = a^2 k (d+k) C_{u,v} + a^2 k d \mu_u \mu_v + a^2 k b_4  \left(1-\frac{1}{d}\right) 
 \left( C_{u,v} + \frac{d}{d-1} \mu_u \mu_v\right)
 \, .
\end{align}

Therefore, the scaling preserving the mean covariance for centered time series with $\mu_u=0=\mu_v$ is then given by 
\begin{align}
 a= \frac{1}{c_2 \sqrt{k[d+k+b_4 (1-1/d)]}}\, .
\end{align}
Thus the main effect of changing the distribution is a fairly obvious scaling which corresponds to normalization of the variance of $B$ to one.
The effect of the fourth cumulant is insignificant, unless it is at least as large as $k$ and $d$.  The third cumulant plays no role in the above computation; it will, however, affect the value of the variance of $C_{Pu,Pv}$.  In fact, 
there are quite a few new terms introduced to the computation of $\Var(C_{Pu,Pv})$.  All of these, however, are still expected to be subdominant to the Gaussian contribution, as long as the higher order cumulants are not comparable to $k$ and $d$. As in the explicit example above, each higher order cumulant should merely introduce new restrictions reducing the combinatorial factors arising from the pairing partitions computed in the proof of the Theorem.
\end{remark}

\end{document}